\documentclass{aa}  
\usepackage[dvipsnames]{xcolor}

\usepackage{longtable, booktabs}
\usepackage[breaklinks, colorlinks, citecolor=blue]{hyperref}

\usepackage{graphicx}
\usepackage{txfonts}
\begin{document}

   \title{Constraining the mass and redshift evolution of the hydrostatic mass bias using the gas mass fraction in galaxy clusters}

   \author{R. Wicker
  \inst{1}
  ,
  M.Douspis\inst{1}, L. Salvati\inst{1}, N. Aghanim \inst{1}
  }

   \institute{Université Paris-Saclay, CNRS, Institut d’Astrophysique Spatiale, 91405, Orsay, France\\
      \email{raphael.wicker@ias.u-psud.fr}
     }

   \date{Received XXX; accepted XXX}

 
  \abstract
  {The gas mass fraction in galaxy clusters is a convenient probe to use in cosmological studies, as it can help derive constraints on a range of cosmological parameters.
  This quantity is, however, subject to various effects from the baryonic physics inside galaxy clusters, which may bias the obtained cosmological constraints. 
  Among different aspects of the baryonic physics at work, in this paper we focus on the impact of the hydrostatic equilibrium assumption.
  We analyzed the hydrostatic mass bias $B$, constraining a possible mass and redshift evolution for this quantity and its impact on the cosmological constraints.
  To that end, we considered cluster observations of the {\it Planck}-ESZ sample and evaluated the gas mass fraction using X-ray counterpart observations.
  We show a degeneracy between the redshift dependence of the bias and cosmological parameters.
  In particular we find evidence at $3.8 \sigma$ for a redshift dependence of the bias when assuming a {\it Planck} prior on $\Omega_m$. 
  On the other hand, assuming a constant mass bias would lead to the extremely large value of $\Omega_m > 0.860$. 
  We show, however, that our results are entirely dependent on the cluster sample under consideration.
  In particular, the mass and redshift trends that we find for the lowest mass-redshift and highest mass-redshift clusters of our sample are not compatible.
  In addition, we show that assuming self-similarity in our study can  impact the results on the evolution of the bias, especially with regard to the mass evolution.
  Nevertheless, in all the analyses, we find a value for the amplitude of the bias that is consistent with $B \sim 0.8$, as expected from hydrodynamical simulations and local measurements. However, this result is still in tension with the low value of $B \sim 0.6$ derived from the combination of cosmic microwave background primary anisotropies with cluster number counts.
  }

       
    \maketitle
%

\section{Introduction}
Galaxy clusters are the most massive gravitationally bound systems of our Universe.
As such they contain a wealth of cosmological and astrophysical information. 
They can be used either as powerful cosmological probes (\cite{1993Natur.366..429W}, \cite{2011ARA&A..49..409A}) or as astrophysical objects of study to better characterise the physics of the intra-cluster medium (ICM, \cite{1988xrec.book.....S}) and how these structures are connected to the rest of the cosmic web.
Constraints on the matter density, $\Omega_m$, or the amplitude of the matter power spectrum, $\sigma_8$, can be inferred from several galaxy cluster observables.
We could, for instance, use cluster number counts, their clustering, or the properties of their gas content to constrain cosmological parameters (see e.g. \cite{2012ARA&A..50..353K} or \cite{2011ARA&A..49..409A} for reviews). Among the more recent probes, we can also cite the cluster sparsity \citep{10.1093/mnras/stt2050}. 

The baryon budget of these objects is also interesting in terms of the aspect of galaxy clusters as cosmological probes and especially the hot gas of the ICM that composes the major part of the baryonic matter inside clusters \citep{2022A&A...664A.198G}.
Indeed the gas mass fraction of galaxy clusters, $f_{gas}$, is considered to be a good proxy for the universal baryon fraction \citep{2011ASL.....4..204B} and can be used to constrain cosmological parameters including the matter density, $\Omega_m$, the Hubble parameter, $h$, the dark energy density, $\Omega_{DE}$, or the equation of state of dark energy $w$ (see e.g. \cite{2008MNRAS.383..879A}, \cite{2020JCAP...09..053H}, \cite{2022MNRAS.510..131M} and references therein).

The gas content inside galaxy clusters is, however, also affected by baryonic physics. 
Such baryonic effects need to be taken into account while performing the cosmological analysis, as they introduce systematic uncertainties in the final constraints (see e.g. the discussions from \cite{2003ApJ...591..515M}, \cite{2003ApJ...591..526M}, \cite{2007MNRAS.380..437P}, \cite{2012MNRAS.427.1298H}, \cite{2013ApJ...767..116M}, \cite{2013MNRAS.432.3508R}, \cite{2018A&A...620A..78S}).

For instance, feedback mechanisms inside clusters, such as active galactic nuclei (AGN) heating, can drive gas out of the potential wells, resulting in slightly gas-depleted clusters.
This depletion of galaxy clusters' gas with respect to the universal baryon fraction is accounted for by the depletion factor $\Upsilon$ (\cite{1998ApJ...503..569E}, \cite{2007MNRAS.377...41C}).
This depletion factor has been thoroughly studied in hydrodynamical simulations throughout the years, in clusters (\cite{2005ApJ...625..588K}, \cite{2013MNRAS.431.1487P}, \cite{2016MNRAS.457.4063S}, \cite{2020MNRAS.498.2114H}, and references therein) as well as in filaments \citep{2022A&A...661A.115G}. 
As a result, this parameter can be very well constrained and robustly predicted in numerical simulations.
Secondly, galaxy clusters are often assumed to be in hydrostatic equilibrium (HE hereafter). Nevertheless, non-thermal processes such as turbulence, bulk motions, magnetic fields or cosmic rays (\cite{2009ApJ...705.1129L}, \cite{2009A&A...504...33V}, \cite{2012ApJ...758...74B}, \cite{2014ApJ...792...25N}, \cite{2015MNRAS.448.1020S}, \cite{2016ApJ...827..112B}), might cause a departure from the equilibrium condition in the ICM. 
Therefore, the HE assumption leads to cluster mass estimations biased toward lower values with respect to the total cluster mass. 
The impact of non-thermal processes, and therefore an evaluation for the bias in the cluster mass estimation, was first considered in hydrodynamical simulations \citep{2006MNRAS.369.2013R}.
Since then, a parametrization for this mass bias has been introduced also in observations, for instance when detecting clusters in X-ray or millimeter wavelengths, the latter exploiting the thermal Sunyaev-Zeldovich effect (\cite{1972CoASP...4..173S}, tSZ hereafter) (see \cite{2019SSRv..215...25P} for a review). We stress that this bias affects all the observables which might assume hydrostatic equilibrium when evaluating cluster masses and, thus, the gas mass fraction. Throughout the paper we define this mass bias as $B=M_{HE}/M_{tot}$.

If the depletion factor is well constrained and understood, this is not the case for the hydrostatic mass bias as its value is still under debate.
In a number of analyses including weak lensing works (Weighing the Giants (WtG) \citep{2014MNRAS.443.1973V}, the Canadian Cluster Comparison Project (CCCP) \citep{2015MNRAS.449..685H}, \cite{2016MNRAS.461.3794O}, \cite{2017MNRAS.472.1946S}), tSZ number counts \citep{2019A&A...626A..27S}, X-ray observations \citep{2019A&A...621A..40E}, or hydrodynamical simulations (\cite{2017MNRAS.465.2936M}, \cite{2022MNRAS.514..313B}), this mass bias is estimated to be around $B \sim 0.8-0.85$. 
The works from \cite{2014A&A...571A..20P}, \cite{2016A&A...594A..24P} however show that to alleviate the tension on the amplitude of the matter power spectrum, $\sigma_8$, between local and cosmic microwave background (CMB hereafter) measurements, a much lower $B$ is needed. 
Indeed, when combining CMB primary anisotropies and tSZ cluster counts, CMB measurements drive the constraining power on the cosmological parameters and, thus, on the bias, favoring a bias $B \sim 0.6-0.65$.
This result was confirmed later on by \cite{2018A&A...614A..13S} and \cite{2020A&A...641A...6P}.
In addition, the study from \cite{2018A&A...614A..13S} shows that when forcing $B = 0.8$ while assuming a {\it Planck} cosmology, the observed cluster number counts are way below the number counts predicted using the CMB best-fit cosmological parameters.
In other words, when assuming $B = 0.8$ with a CMB cosmology, one predicts approximately thrice as many clusters as what is actually observed.

As the precise value of the mass bias is still an open matter and has a direct impact on the accuracy and precision  of the cosmological constraints deduced from galaxy clusters, we propose a new and independent measurement of this quantity. 
In this paper we use the gas mass fractions of 120 galaxy clusters from the {\it Planck}-ESZ sample \cite{2011A&A...536A...8P} to bring robust constraints on the value of the hydrostatic bias.
More importantly, we aim at studying the potentiality of variations of the bias with mass and redshift.
Such studies on mass and redshift trends of $B$ have already been carried out in works using weak lensing (\cite{2014MNRAS.443.1973V}, \cite{2015MNRAS.449..685H}, \cite{2016MNRAS.456L..74S}, \cite{2017MNRAS.468.3322S}) or tSZ number counts \citep{2019A&A...626A..27S}, sometimes giving contradictory results.
In this work we measure and use gas mass fractions from a sample to get new independent constraints on $B$ as well as on its mass and redshift evolution.
This study is also aimed at investigating the role that an evolution of the bias would have on the cosmological constraints we obtain from $f_{gas}$ data.

After describing the theoretical modelling and our cluster sample in Section \ref{sect:theoretical_model_data}, we detail our methods for the data analysis in Section \ref{sect:methods}.
We show our results in Section \ref{sect:results}, first focussing on the effect of assuming a varying bias on our derived cosmological constraints, then looking at the sample dependence of our results.
Finally, we discuss our results in Section \ref{sect:discussion} and draw our conclusions in Section \ref{sect:conclusions}.
Throughout the paper, we assume a reference cosmology with $\mathrm{H_0 = 70 km.s^{-1}.Mpc^{-1}}$, $\Omega_m = 0.3$ and $\Omega_\Lambda = 0.7$.

\section{Theoretical modelling and data}
\label{sect:theoretical_model_data}

\begin{figure}
\centering
    \includegraphics[width=0.5\textwidth, trim = {0.5cm 3.5cm 0.25cm 4.5cm}, clip]{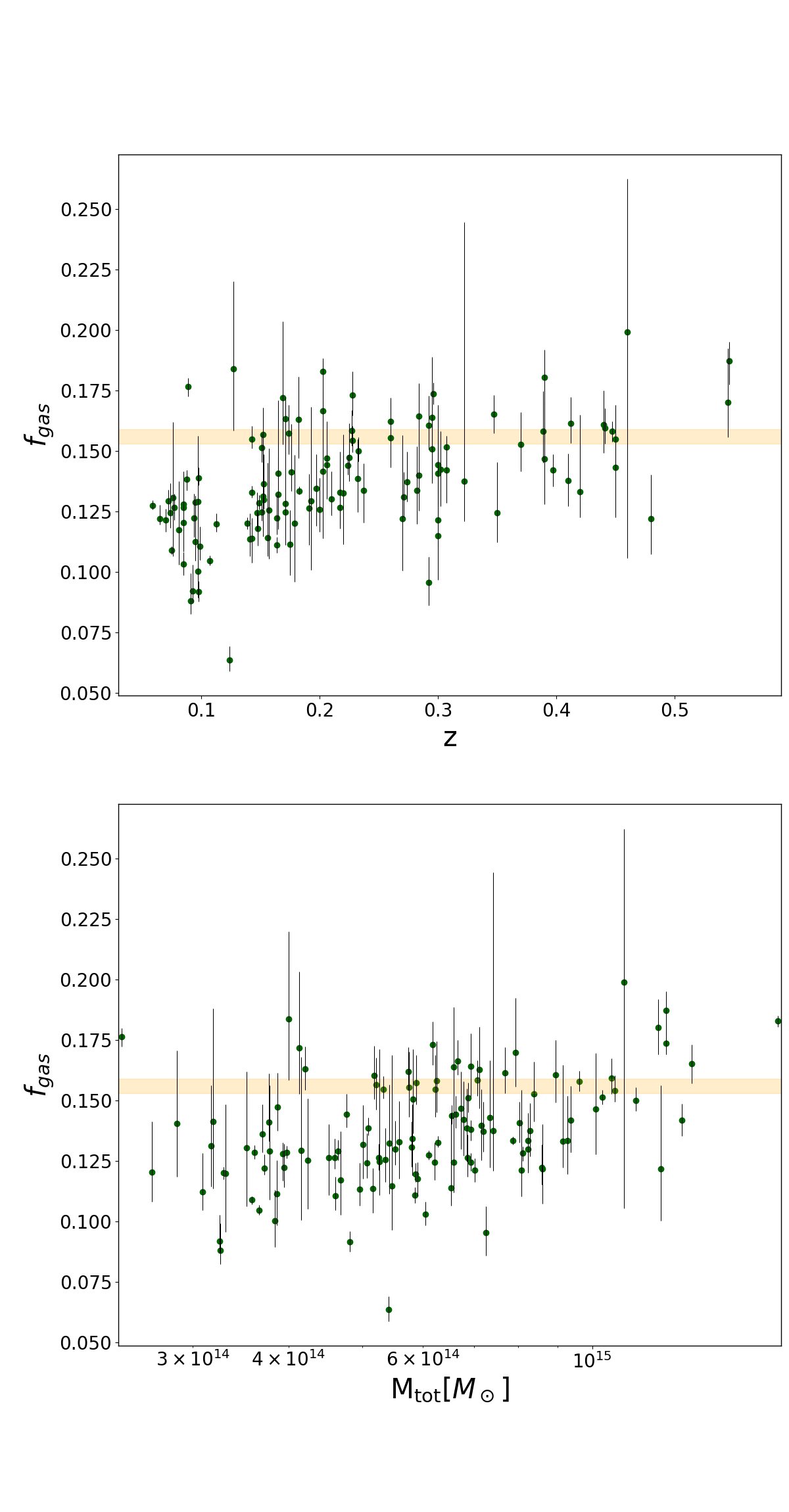}
    \caption{Gas mass fraction of {\it Planck}-ESZ clusters with respect to redshift (top) and with respect to cluster mass (bottom). The yellow bands in both plots mark the \cite{2020A&A...641A...6P} value of $\Omega_b/\Omega_m = 0.156 \pm 0.003$.}
\label{fig:fgas_sample}
\end{figure}

\subsection{Gas fraction sample}
The gas mass fraction is defined as the ratio of the gas mass over the total mass of the cluster :
\begin{equation}
\label{eq:fgas_def}
    f_{gas} = \frac{M_{gas}}{M_{tot}}.
\end{equation}
Using X-ray observations, the gas mass is obtained by integrating the density profile $\rho(r)$ inside a certain radius, $r$, as shown in Equation \ref{eq:gas_mass} below
\begin{equation}
    \label{eq:gas_mass}
    M_{gas}(<r) = \int^r_0 4\pi r'^2 \rho(r')dr'.
\end{equation}
Here, the density profile $\rho(r)$ is obtained from the electron density profile $n_e(r)$, measured from X-ray observations. We have :
\begin{equation}
    \rho(r) = \mu m_p (n_e(r) + n_p(r)),
\end{equation}
where $\mu$ is the mean molecular weight, $m_p$ the proton mass, $n_e$ the electron number density, and $n_p$ the proton number density with $n_e = 1.17 n_p$ in a fully ionised gas.
Using the density profile and a temperature profile, $T(r)$, the total hydrostatic mass, $M_{HE}$, can be computed by solving the hydrostatic equilibrium equation shown below :
\begin{equation}
    \label{eq:hydro_mass}
        M_{HE}(<r) = - \frac{rk_B T(r)}{G \mu m_p} \left( \frac{\mathrm{d \ln} \rho(r)}{\mathrm{d \ln} r} + \frac{\mathrm{d \ln} T(r)}{\mathrm{d \ln} r} \right),
\end{equation}
where $k_B$ is the Boltzmann constant and $G$ is the gravitational constant. 

Similarly to the gas mass and the hydrostatic mass, the cluster gas mass fraction is evaluated within a characteristic radius, determined by the radius of the mass measurements.
We define this radius, $R_{\Delta}$, as the radius that encloses $\Delta$ times the critical density of the Universe, $\rho_c = 3 H^2/8 \pi G$. 
The gas content inside galaxy clusters is affected by baryonic physics and the impact of the different astrophysical processes might depend on the considered cluster radius, $R_{\Delta}$. 
In this work, we focus on gas fractions taken at $R_{500}$, and from now on all the quantities we consider are taken at $R_{500}$.
We note that this radius is larger than most studies using the gas fraction of galaxy clusters as a cosmological probe, which are generally carried out at $R_{2500}$ (\cite{2002MNRAS.334L..11A}, \cite{2008MNRAS.383..879A}, \cite{2011ARA&A..49..409A}, \cite{2014MNRAS.440.2077M}, \cite{2020JCAP...09..053H}, \cite{2022MNRAS.510..131M}). 
We briefly discuss this choice of radius in Section \ref{sect:comparison_R2500}.

We computed the gas fraction for the clusters in the {\it Planck}-ESZ survey \citep{2011A&A...536A...8P}.
In particular, we considered 120 out of 189 clusters of the {\it Planck}-ESZ sample, for which we have follow-up X-ray observations by XMM-{\it Newton} up to $R_{500}$ (here called the ESZ sample for simplicity). Our sample therefore spans a total mass range from $2.22 \times 10^{14} \mathrm{M_\odot}$ to $1.75 \times 10^{15} \mathrm{M_\odot}$ and redshift range from 0.059 to 0.546. 
We started from the gas and total masses of the clusters derived in \cite{2020ApJ...892..102L}. We refer to their work for the detailed analysis of the mass evaluation. We just stress here that the total cluster masses were obtained assuming hydrostatic equilibrium, as shown in Equation \ref{eq:hydro_mass}, thus inducing the presence of the hydrostatic mass
bias. From these gas masses and hydrostatic masses, we computed the gas fraction following Equation \ref{eq:fgas_def} to find that they are within the range $[0.06, 0.20]$. 
We note that there are correlations between the gas mass and hydrostatic mass measurements. 
These correlations induce a corrective term when computing the error bars of $f_{gas}$.
We check that this correlation term (provided by Lorenzo Lovisari, private communication) introduces a negligible contribution to the total error budget.
We propagate the asymmetric errors from the mass measurements to obtain the uncertainty on $f_{gas}$.
We also note that the Malmquist bias may affect our final results; however, the discussions in \cite{2020ApJ...892..102L} and \cite{2021ApJ...914...58A} show that this effect is negligible in the case of this sample.

We give the redshifts, gas masses, hydrostatic masses and gas fractions of the clusters from our sample in Table \ref{tab:cluster_sample} in the appendix.
We show in Figure \ref{fig:fgas_sample} the observed gas fraction in this sample, with respect to redshift and mass.
We also compare these $f_{gas}$ values to the universal baryon fraction $\Omega_b/\Omega_m = 0.156\pm 0.03$ from {\it Planck} 2018 results \citep{2020A&A...641A...6P}.

\subsection{Modelling of the observed gas fraction}
The hydrostatic mass is used to evaluate the total cluster mass, yet it is biased low with respect to the true total mass by a factor $B=M_{HE}/M_{true}$. 
The measured gas mass fraction is thus :
\begin{equation}
    \label{eq:hydro_fgas}
    f_{gas, mes} = \frac{M_{gas}}{M_{HE}} = \frac{M_{gas}}{B \times M_{true}} = \frac{1}{B} \times f_{gas, true}.
\end{equation}
Besides the hydrostatic mass bias, the measured gas fraction depends on a variety of instrumental, astrophysical, and cosmological effects.
One way of quantifying these effects is to compare the gas fraction we obtain from gas mass and hydrostatic mass measurements to the theoretical (hydrostatic) gas fraction we expect from Equation \ref{eq:fgas_model} below, from \cite{2008MNRAS.383..879A}. In this equation (and in the rest of the paper), the quantity noted as $X^{ref}$ is the quantity $X$ in our reference cosmology :
\begin{equation}
    f_{gas, Th}(M, z) = K \frac{\Upsilon(M,z)}{B(M,z)} A(z) \left( \frac{\Omega_b}{\Omega_m}\right) \left( \frac{D_A^{ref}(z)}{D_A(z)}\right)^{3/2} - f_*.
    \label{eq:fgas_model}
\end{equation}
Here, $K$ is an instrumental calibration correction. 
We take $ K = 1 \pm 0.1$ from \cite{2008MNRAS.383..879A} and discuss the soundness of this assumption in Section \ref{sect:instrumental_calibration}.
Regarding the astrophysical contributions, $\Upsilon(M,z)$ is the baryon depletion factor, describing how baryons in clusters are depleted with respect to the universal baryon fraction and $B(M,z)$ is the hydrostatic mass bias we discuss above.
Finally $\Omega_b/\Omega_m$ is the universal baryon fraction, $D_A$ is the angular diameter distance, $f_*$ is the stellar fraction, and $A(z)$ is an angular correction which we show in Equation \ref{eq:A_z}:

\begin{equation}
    \label{eq:A_z}
    A(z) = \left (\frac{\theta_{500}^{ref}}{\theta_{500}} \right)^\eta \simeq \left( \frac{H(z)D_A(z)}{\left [H(z)D_A(z) \right]^{ref}} \right)^\eta.
\end{equation}
The parameter $\eta$ accounts for the slope of the $f_{gas}$ profiles enclosed in a spherical shell. Here we take $\eta = 0.442$ from \cite{2014MNRAS.440.2077M}. With
$A(z)$ being, however, very close to one for realistic models and within our range of parameters, the value of this parameter has a negligible impact on our results.

We note that this model is valid as long we are assuming self-similarity, which we do here. 
Deviations from self-similarity may induce supplementary dependencies. 
We thus discuss our hypothesis in Section \ref{sect:contributions_to_fgas} and Appendix \ref{annex:B}.

\section{Methods}
\label{sect:methods}

\begin{table*}[h!]
    \centering
    \caption{Set of priors used in our analysis.}
    \begin{tabular}{cccc}
    \hline \hline
     & {\bf Bias evolution study} & {\bf Sample dependence of the results}  & {\bf Reference}\\
    \hline
    Parameter & Prior & Prior & \\
    \hline
    $B_0$  &  --  &  $\mathcal{U}(0.3, 1.7)$ & -- \\
    $B(z_{CCCP}, M_{CCCP})$ & $\mathcal{N}(0.84, 0.04)$ & -- & 1 \\
    $f_*$ & $\mathcal{N}(0.015, 0.005)$ & $\mathcal{N}(0.015, 0.005)$ & 2\\
    $\Upsilon_0 $ & $\mathcal{N}(0.85, 0.03)$ & $\mathcal{N}(0.85, 0.03)$ & 3\\
    $K$ & $\mathcal{N}(1, 0.1)$ & $\mathcal{N}(1, 0.1)$ & 4 \\
    $\sigma_f$ & $\mathcal{U}(0, 1)$ & $\mathcal{U}(0, 1)$ & -- \\
    $h$ & $\mathcal{N}(0.674, 0.005)$ & $\mathcal{N}(0.674, 0.005)$ & 5\\
    $\Omega_b/\Omega_m$ & $\mathcal{U}(0.05, 0.3)$ & $\mathcal{N}(0.156, 0.003)$ & 5\\
    $\Omega_m$ & $\mathcal{U}(0.01, 1)$ {\bf (CB, VB)} or $\mathcal{N}(0.315, 0.007)$ {\bf (VB + $\Omega_m$)} & $\mathcal{N}(0.315, 0.007)$ & 5\\
    \hline
    $\alpha$ & Fixed at 0 {\bf (CB)} or $\mathcal{U}(-2, 2)$ {\bf (VB, VB + $\Omega_m$)} & $\mathcal{U}(-2, 2)$ & -- \\
    $\beta$ & Fixed at 0 {\bf (CB)} or $\mathcal{U}(-2, 2)$ {\bf (VB, VB + $\Omega_m$)} & $\mathcal{U}(-2, 2)$ & -- \\
    \hline \hline
    \end{tabular}
    \tablebib{(1)~\citet{2020MNRAS.497.4684H}; (2)\citet{2019A&A...621A..40E}; (3)\citet{2013MNRAS.431.1487P}; (4)\citet{2008MNRAS.383..879A};     (5)\citet{2020A&A...641A...6P}.}
    \tablefoot{A prior noted $\mathcal{U}(l, u)$ is a uniform prior of lower bound $l$ and upper bound $u$, while a prior noted $\mathcal{N}(\mu, \sigma)$ is a gaussian prior of mean $\mu$ and standard deviation $\sigma$.}
    \label{tab:priors}
\end{table*}
\begin{figure}
    \centering
    \includegraphics[width = 0.5\textwidth, trim = {3cm 3cm 1cm 7cm}, clip]{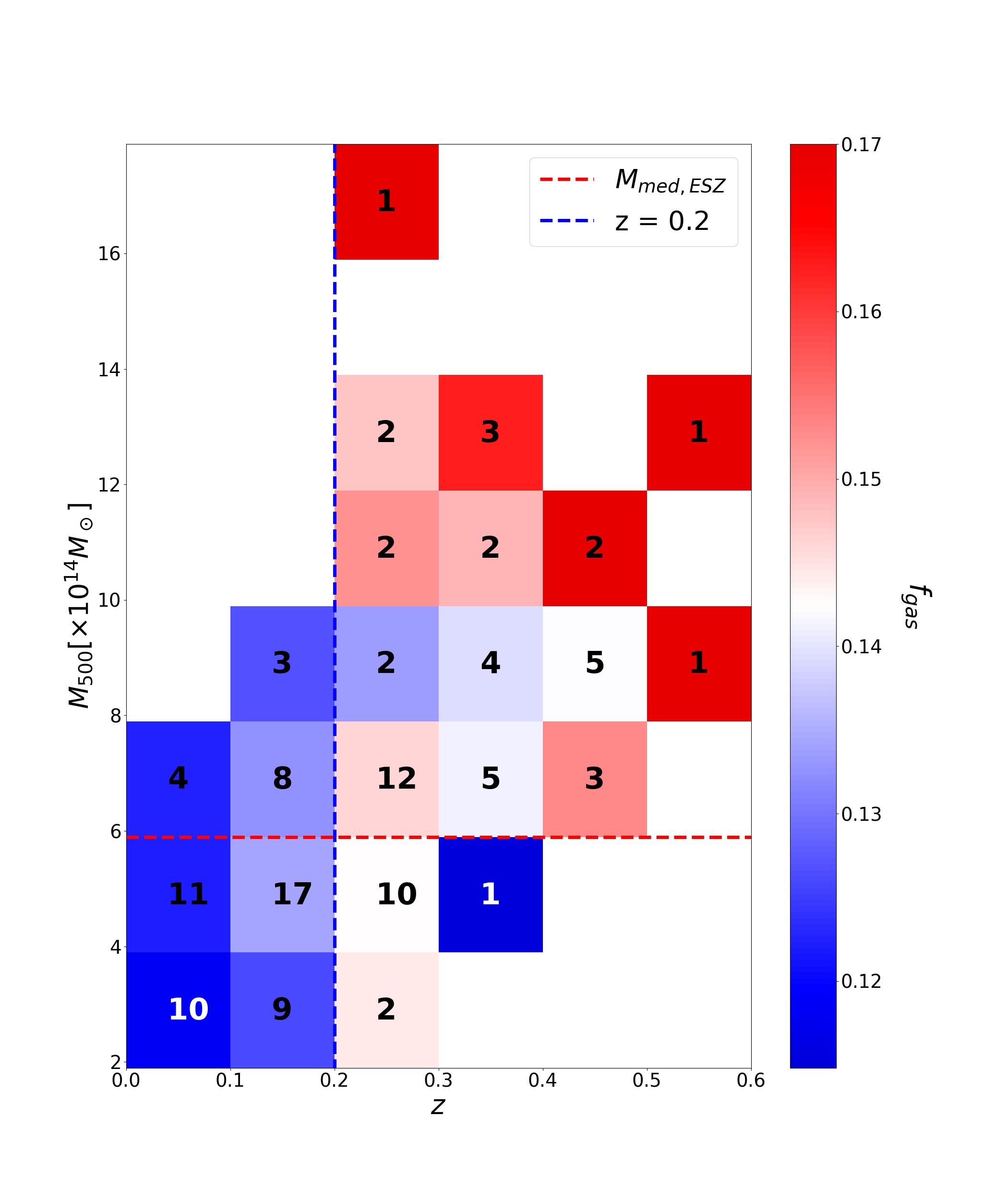}
    \caption{Binned mass-redshift plane of the {\it Planck}-ESZ sample. Inside each bin we compute the mean value of $f_{gas}$. We show the number of clusters included in each bin and mark the delimitation of each subsample.}
    \label{fig:binned_Mz}
\end{figure}
Our purpose in this work is to use the gas mass fraction of galaxy clusters to constrain the value of the hydrostatic mass bias and, in particular, its evolution with mass and redshift.
We also want to study the role of such an evolution of the bias on the subsequent cosmological constraints derived from $f_{gas}$ data.
We recall here that all constraints derived from gas mass fraction data are obtained by comparing the measured gas fraction to the theoretical gas fraction expected from Equation \ref{eq:fgas_model}, which is proportional to the universal baryon fraction.
Besides this proportionality, all the other constraints are deduced as corrections needed to match the observed $f_{gas}$ in clusters to the constant $\Omega_b/\Omega_m$.
As discussed in Section \ref{sect:theoretical_model_data}, one of these correction terms accounts for the baryonic effects taking place in the ICM.
These baryonic effects are accounted for by the depletion factor $\Upsilon(M,z)$ and the hydrostatic bias $B(M, z)$, which we are interested in.
As shown in Equation \ref{eq:fgas_model}, we cannot constrain independently $B(M, z)$ and $\Upsilon(M, z)$, as the two parameters are strongly degenerated.
What we have access to instead is the ratio of the two quantities, $\Upsilon(M, z)/B(M, z)$.
In order to break this degeneracy and properly constrain the bias, strong constraints on the depletion factor and its evolution with mass and redshift are required.
Obtaining such results is however out of the scope of this paper, and we retrieve these constraints from hydrodynamical simulations works. 

The depletion factor is known to vary with mass, however this evolution is particularly strong for groups and low mass clusters ($<2.10^{14} M_\odot$) while it becomes negligible at the high masses we consider (see the discussion in Section 3.1.1 of \cite{2019A&A...621A..40E} based on results from The Three Hundred Project simulations \citep{2018MNRAS.480.2898C}).
Works from \cite{2013MNRAS.431.1487P} and \cite{2013ApJ...777..123B} also show that the depletion factor is constant with redshift when working at $R_{500}$.
Throughout the paper, we thus assume a constant depletion factor with mass and redshift $\Upsilon_0 = 0.85 \pm 0.03$, based on hydrodynamical simulations from \cite{2013MNRAS.431.1487P}.

\subsection{Bias evolution modelling}
In order to analyze a possible mass and redshift evolution of the mass bias, we consider a power law evolution for the hydrostatic bias, with pivot masses and redshifts set at the mean values of the considered cluster sample :
\begin{equation}
    B(z, M) = B_0 \left( \frac{M}{\left< M \right>}\right)^\alpha \left( \frac{1+z}{\left< 1+z \right>}\right)^\beta 
    \label{eq:b_powerlaw}
\end{equation}
with $B_0$ as the amplitude.

We chose this model for the sake of simplicity following what is done in \cite{2019A&A...626A..27S}, as the exact dependence of $B$ on mass and redshift is not known.
In Section \ref{sect:parametrization} we discuss the role of this parametrization by comparing our results with a linear evolution of $B$, and find results similar to those of the power law description.
The complete likelihood function thus writes, assuming no deviation from self-similarity : \\ \\
\begin{strip}
\large
\begin{equation}
\label{eq:likelihood}
- 2 \ln \mathcal{L} = \sum_i \left (\ln{2\pi s_i^2} + \frac{\left(f_{gas, i} - K \frac{\Upsilon_0}{B_0} \left( \frac{M_i}{\left< M \right>}\right)^{-\alpha} \left( \frac{1+z_i}{\left< 1+z \right>}\right)^{-\beta} A(z_i, h, \Omega_m) \frac{\Omega_b}{\Omega_m} \left( \frac{D_A^{ref}(z_i)}{D_A(z_i, h, \Omega_m)}\right)^{3/2} + f_* \right)^2}{s_i^2}\right),
\end{equation}
\end{strip}
where 
\begin{equation}
    s_i^2 = \sigma_i^2 + f_{gas, Th}^2 \sigma_f^2,
\end{equation}
with $\sigma_i$ as the uncertainties on the gas fraction data and $f_{gas, Th}$ as the gas fraction expected from Equation \ref{eq:fgas_model}. 
The parameter $\sigma_f$ is the intrinsic scatter of the data, which we treat as a free parameter.

\subsection{Free cosmology study}
In the first part of our analysis we simply assess the necessity to consider an evolving bias, by looking at the impact of assuming such a variation on the subsequent cosmological constraints.
To do so, we compare the posterior distributions in three different cases :
In the first case we let free the baryon fraction, $\Omega_b/\Omega_m$, the matter density, $\Omega_m$, and the parameters accounting for the variation of the bias $\alpha$, $\beta$, and $B_0$. 
 We refer to this scenario as 'VB'.
In the second case we let free only the cosmological parameters and fix the bias parameters to $\alpha=0$ and $\beta=0$, resulting in a constant bias at the value $B_0$.
This scenario will be noted in the rest of the paper as 'CB'.
Due to a degeneracy between $\beta$ and $\Omega_m$ which we discuss later on, we also look at our results when leaving the set of parameters ($B_0, \alpha, \beta, \Omega_b/\Omega_m$) free but assuming a prior on $\Omega_m$, in order to break this degeneracy and constrain $\beta$ more accurately.
We call this scenario 'VB + $\Omega_m$'.

The set of parameters for which we assume flat priors in this part of the study is thus ($B_0, \alpha, \beta, \Omega_b/\Omega_m, \Omega_m, \sigma_f$), as we show in the first column of Table \ref{tab:priors}.
This work is performed on the entirety of our cluster sample, for which our mean mass and redshift are :
\begin{equation}
    (M_{full}, z_{full}) = (6.42 \times 10^{14} M_\odot, 0.218)
    \label{eq:pivots_full_ESZ}
\end{equation}
Throughout this whole part of the analysis, we consider a prior on the total value of the bias for a certain cluster mass and redshift, taken from the Canadian Cluster Comparison Project - Multi Epoch Nearby Cluster Survey (CCCP-MENeaCS) analysis \cite{2020MNRAS.497.4684H} :
\begin{equation*}
    B(z_{CCCP}, M_{CCCP}) = 0.84 \pm 0.04,
\end{equation*}
where $z_{CCCP} = 0.189$ and $M_{CCCP} = 6.24 \times 10^{14}M_\odot$ are the mean mass and redshift for the CCCP-MENeaCS sample.
\subsection{Sample dependence tests}
In the second part, we looked into possible sample dependencies of our results regarding the value of the bias parameters $B_0$, $\alpha$, and $\beta$.
To that end, we focus on different subsamples within the main sample, based on mass and redshift selections.
Matching the selection from weak lensing studies such as the Comparing Masses in Literature ({\sc CoMaLit}) \citep{2017MNRAS.468.3322S} or Local Cluster Substructure Survey ({\sc LoCuSS}) \citep{2016MNRAS.456L..74S} studies, we operate a redshift cut at $z = 0.2$, differentiating clusters that are above or below this threshold value.
This choice was also motivated by the results from \cite{2019A&A...626A..27S}, investigating the hydrostatic mass bias from the perspective of tSZ number counts. Their study showed that the trends in the mass bias depended on the considered redshift range, with results changing when considering only clusters with $z>0.2$.
We also performed a mass selection, differentiating between the clusters that are above or below the median mass of the sample $M_{med} = 5.89\times 10^{14}M_\odot$.

In summary, the samples we consider in this study are the following :

The full sample of 120 clusters, with the mean mass and redshift given previously in Equation \ref{eq:pivots_full_ESZ}.
The second subsample is composed of clusters with $z < 0.2$ and $M <  5.89\times 10^{14}M_\odot$. We consider them in the '{\it low Mz}' subsample, which contains 47 clusters. The mean mass and redshift are 
    \begin{equation}
        (M_{lowMz}, z_{lowMz}) = (4.26 \times 10^{14} M_\odot, 0.126)
        \label{eq:pivots_lowMz}
    \end{equation}
Finally our third subsample is constituted of clusters with $z > 0.2$ and $M >  5.89\times 10^{14}M_\odot$. We consider them in the '{\it high Mz}' subsample, which contains 45 clusters. The mean mass and redshift are
    \begin{equation}
        (M_{highMz}, z_{highMz}) = (8.86 \times 10^{14} M_\odot, 0.333)
        \label{eq:pivots_highMz}
    \end{equation}
We did not consider the 'low z - high M' and 'high z - low M' subsamples as by construction they do not contain enough clusters (15 and 13, respectively) to obtain meaningful results.
For illustration purposes, we show in Figure \ref{fig:binned_Mz} the binned mass-redshift plane of our sample, with the delimitation of the different subsamples. 
Inside each bin, we computed the mean value of $f_{gas}$ if at least one clusters is inside the bin.

To carry out this study of the sample dependence, we compared the posterior distributions obtained when running our MCMC on the three aforementioned samples independently. 
We note that we kept the prior on $\Omega_m$ for this part of the study, and that we add a prior on $\Omega_b/\Omega_m$. 
This choice is motivated by the presence of a degeneracy between the baryon fraction and the amplitude of the bias $B_0$.
This degeneracy is broken in the first part of the analysis by assuming a prior on the total value of the bias.
However, when trying to compare all the bias parameters between samples (including the amplitude), we do not wish to be dependent on such a prior.
The universal baryon fraction being well known and constrained, such a prior is a convenient way to break the degeneracy between $\Omega_b/\Omega_m$ and $B_0$ and still obtain meaningful results on the value of all the bias parameters. 
The set of parameters following flat priors in this part of the study is thus ($B_0, \alpha, \beta, \sigma_f$), as we show in the second column of Table \ref{tab:priors}. 

In brief, we adopted the list of priors given in Table \ref{tab:priors} to constrain the parameters used to describe our $f_{gas}$ data. 
We fit our model in Equation \ref{eq:fgas_model} to the $f_{gas}$ data with an MCMC approach using the sampler {\tt emcee} \citep{2013PASP..125..306F}. 
We note that the prior we consider on $f_* = 0.015\pm 0.005$ coming from \cite{2019A&A...621A..40E} has close to no effect on our final results, as this term in Equation \ref{eq:fgas_model} is almost negligible.

\section{Results}
\label{sect:results}
\subsection{Bias evolution study}
\label{sect:bias_evolution}
\begin{figure*}
    \centering
    \includegraphics[width=\textwidth, trim = {0.75cm 1cm 0.5cm 0.5cm}, clip]{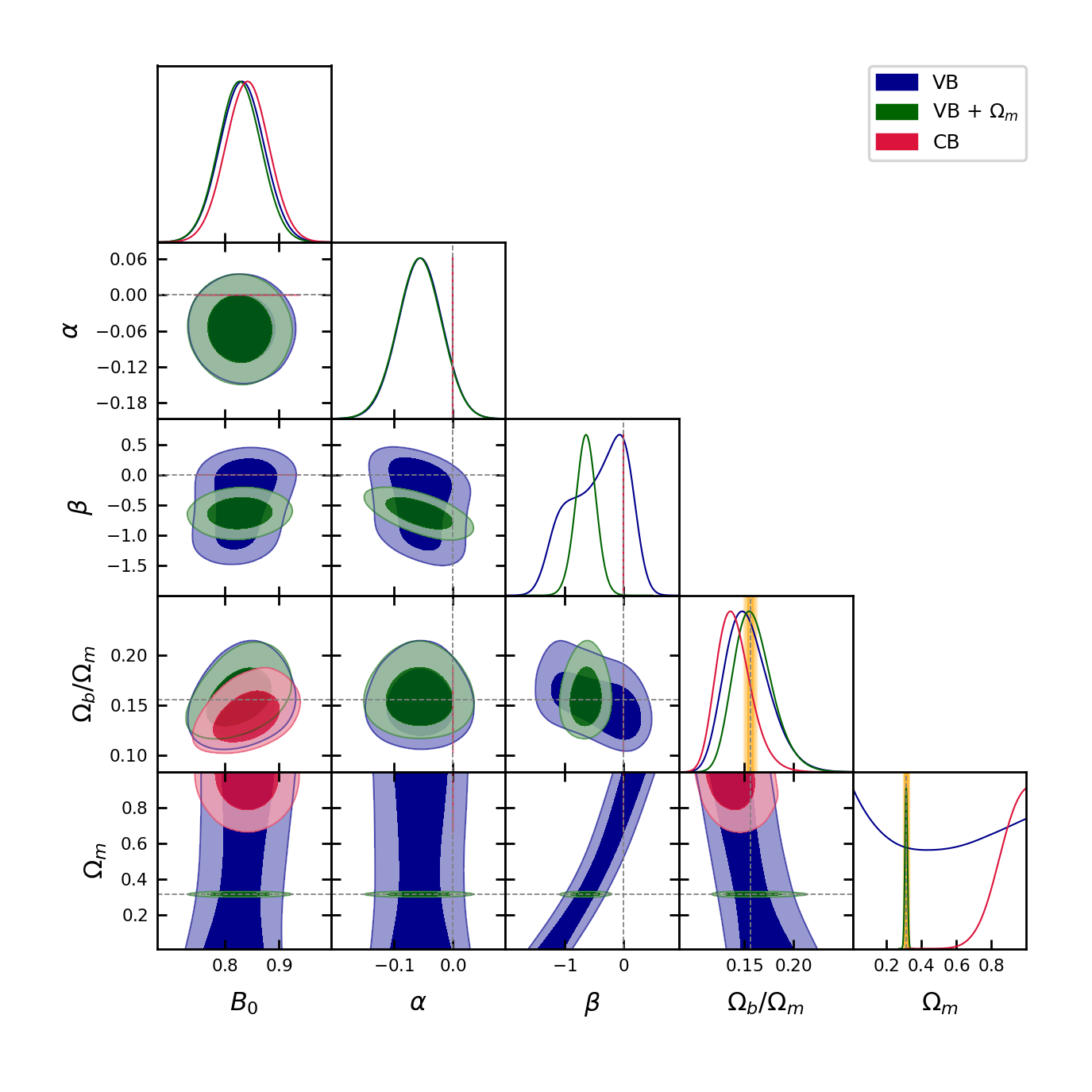}
    \caption{1D and 2D posterior distributions for the CB, VB, and VB + $\Omega_m$ scenarios. The contours mark the 68\% and 95\% confidence level (c.l.). The gray dashed lines highlight reference values for $(\alpha, \beta, \Omega_b/\Omega_m, \Omega_m) = (0, 0, 0.156, 0.315)$. The orange bands mark the \cite{2020A&A...641A...6P} values for $\Omega_b/\Omega_m$ and $\Omega_m$ at 2$\sigma$ c.l.}
    \label{fig:varying_v_constant}
\end{figure*}

\begin{figure}
    \centering
    \includegraphics[width=0.5\textwidth, trim = {0.5cm 0.5cm 0.5cm 0.5cm}, clip]{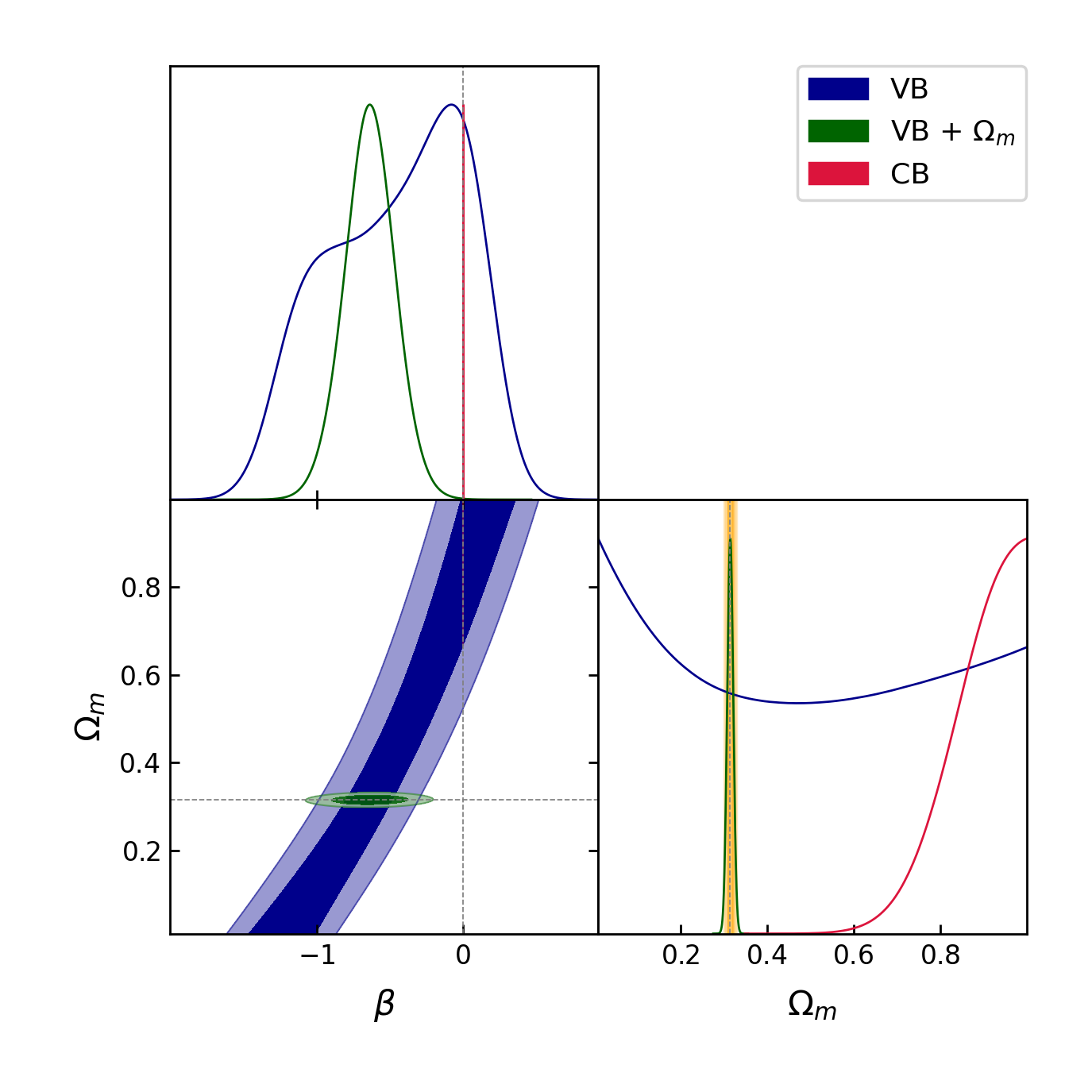}
    \caption{Posterior distributions showing the degeneracy between $\beta$ and $\Omega_m$. The contours mark the 68\% and 95\% c.l. The gray dashed lines highlight reference values for $(\beta, \Omega_m) = (0, 0.315)$. The orange band marks the \cite{2020A&A...641A...6P} value for $\Omega_m$ at 2$\sigma$ c.l.}
    \label{fig:varying_v_constant_degenzoom}
\end{figure}

In the first part of this analysis, we intend on studying the possibility of an evolution of the hydrostatic mass bias with cluster mass and redshift.
To that end, we compare the cosmological constraints obtained on the full sample when considering a constant bias to those obtained when leaving the bias free to vary. Our results are summed up in Figure \ref{fig:varying_v_constant} and in Table \ref{tab:results_varying_v_constant}.

\begin{table*}
    \centering
    \caption{Constraints obtained on bias and cosmological parameters in the CB, VB and VB + $\Omega_m$ scenarios. Uncertainties are given at 68\%c.l.}
    \begin{tabular}{cccc}
    \hline \hline
    {\bf Parameter} & CB & VB & VB + $\Omega_m$ \\
    \hline
    $\mathbf{B_0}$ &  $0.842 \pm 0.040$ & $0.832 \pm 0.041$ & $0.828 \pm 0.039$ \\
    $\boldsymbol{\alpha}$ & 0 & $-0.056\pm 0.037$ & $-0.057\pm 0.038$ \\
    $\boldsymbol{\beta}$ & 0 & $-0.43^{+0.61}_{-0.37}$ & $-0.64 \pm 0.18$ \\
    $\boldsymbol{\Omega_b/\Omega_m}$ & $0.140^{+0.014}_{-0.020}$ & $0.154^{+0.018}_{-0.026}$ &  $0.160^{+0.016}_{-0.025}$ \\
    $\boldsymbol{\Omega_m}$ & $> 0.860$ & -- & $0.315\pm 0.007$ \\
    \hline
    \end{tabular}
    \label{tab:results_varying_v_constant}
\end{table*}

In the VB case, namely when leaving the set of parameters $(B_0, \alpha, \beta, \Omega_b/\Omega_m, \Omega_m, \sigma_f)$ free with a prior on the total value of $B$, we show that a mass-independent bias seems to be favored.
Indeed, with $\alpha  = -0.056\pm 0.037$ we remain compatible with 0 within 2$\sigma$, even though the peak is slightly lower.
We also show that our derived $\Omega_b/\Omega_m$ is fully compatible with the \cite{2020A&A...641A...6P} value, since we obtained $0.154^{+0.018}_{-0.026}$.
We cannot, however, bring such constraints on $\beta$ and $\Omega_m$, as these two parameters are heavily degenerated. 
We zoom in on this degeneracy in Figure \ref{fig:varying_v_constant_degenzoom} and show that higher values of $\beta$ call for higher values of $\Omega_m$.
This degeneracy can be explained by the fact that both parameters entail a redshift dependence on the part of the gas fraction.
Indeed, $\Omega_m$ intervenes in the computation of $D_A(z)$ and $H(z)$, which entirely drive the sensitivity of $f_{gas}$ to cosmology.
We thus argue that this degeneracy between $\beta$ and $\Omega_m$ (which we show here in a simple Flat-$\Lambda \mathrm{CDM}$ model) could also be observed between $\beta$ and any other cosmological parameter, as long as they appear in the computation of $D_A(z)$ or $H(z)$.
As a side note, we show a slight degeneracy between the baryon fraction $\Omega_b/\Omega_m$ and $\beta$, with a lower $\Omega_b/\Omega_m$ implying a higher, closer to 0 $\beta$.
This degeneracy is caused by the combined effect of the degeneracy between $\beta$ and $\Omega_m$, and the degeneracy between $\Omega_b/\Omega_m$ and $\Omega_m$, which is expected (shown in Figure \ref{fig:varying_v_constant}).

As the matter density, $\Omega_m$, has been strongly constrained in a number of works, we chose to assume the \cite{2020A&A...641A...6P} prior shown in Table \ref{tab:priors} on this parameter to break its degeneracy with $\beta$, in the VB + $\Omega_m$ scenario.
The effect of this prior is negligible on the constraints on $\alpha$ and $\Omega_b/\Omega_m$, as we obtain 
$\alpha = -0.057\pm 0.038$ and $0.160^{+0.016}_{-0.025}$, fully compatible with the results obtained without this prior.
On the other hand, the use of this prior allows us to constrain $\beta$. 
We show that the hydrostatic bias seems to show a strong redshift dependence, with $\beta = -0.64\pm 0.18$.
This value of $\beta < 0$ would mean a value of $B$ decreasing with redshift, that is to say, more and more biased masses toward higher redshifts.
We also note a slight degeneracy between $\alpha$ and $\beta$, but this is most probably due to selection effects of the sample.
As the higher redshift clusters tend to mostly have higher masses (see Figure \ref{fig:binned_Mz}), a redshift trend of the bias could then be interpreted as a slight mass trend, explaining this degeneracy.
These selection effects causing the degeneracy are those we explore in the second part of the analysis when looking at the sample dependence of the results (see Sect \ref{sect:sample_dependence}).

The effect of assuming a constant bias ($\alpha = \beta = 0$) in the CB case does not have a strong impact on our constraints on $\Omega_b/\Omega_m$, which peaks just slightly below the {\it Planck}; it does remain compatible, with $\Omega_b/\Omega_m = 0.140^{+0.014}_{-0.020}$.
$B_0$ is slightly above yet compatible with the value found in the varying bias case, with $B_0 = 0.842 \pm 0.040$. This is simply caused by the fact that for a constant bias the amplitude $B_0$ is now completely determined by the total value of the bias $B(z_{CCCP}, M_{CCCP})$.
On the other hand, due to the degeneracy between $\beta$ and $\Omega_m$, imposing no redshift evolution of the bias requires a very high matter density, resulting in $\Omega_m > 0.860$, fully incompatible with the {\it Planck} value.
Such a high matter density is not expected in current standard cosmology and is totally aberrant. 
We note that these results do not completely rely on our prior on the total value of the bias. 
Indeed, we also performed our analysis while assuming a prior from the first results of the CCCP analysis \citep{2015MNRAS.449..685H}:
\begin{equation}
    B(\left<z\right>, \left<M\right>) = 0.780 \pm 0.092
\end{equation}
with $(\left<z\right>, \left<M\right>) = (0.246, 14.83\times 10^{14}h^{-1}M_\odot)$.
Our constraints when considering this prior do not change with respect to $B_(z_{CCCP}, M_{CCCP}) = 0.84 \pm 0.04$, except a small shift on $\Omega_b/\Omega_m = 0.132_{-0.024}^{+0.019}$ in the CB case.

As such, we show that we need to assume an evolution of the hydrostatic mass bias, at least in redshift, to properly describe our $f_{gas}$ data.
In the rest of our study we focus on exploring the bias evolution.
We thus consider only the VB + $\Omega_m$ scenario, to be able to constrain $\beta$ despite its degeneracy with the matter density. 
We also assume a \cite{2020A&A...641A...6P} prior on $\Omega_b/\Omega_m$, as the universal baryon fraction is degenerated with the total value of the bias (see Equation \ref{eq:fgas_model}).
The bias parameters $(B_0, \alpha, \beta)$ are thus left free and we chose not to use a prior on the total value of $B$ going forward. 

\subsection{Sample dependence of the results}
\label{sect:sample_dependence}
\begin{figure}
    \centering
    \includegraphics[width=0.5\textwidth, trim = {0.75cm 0.75cm 0.5cm 0.5cm}, clip]{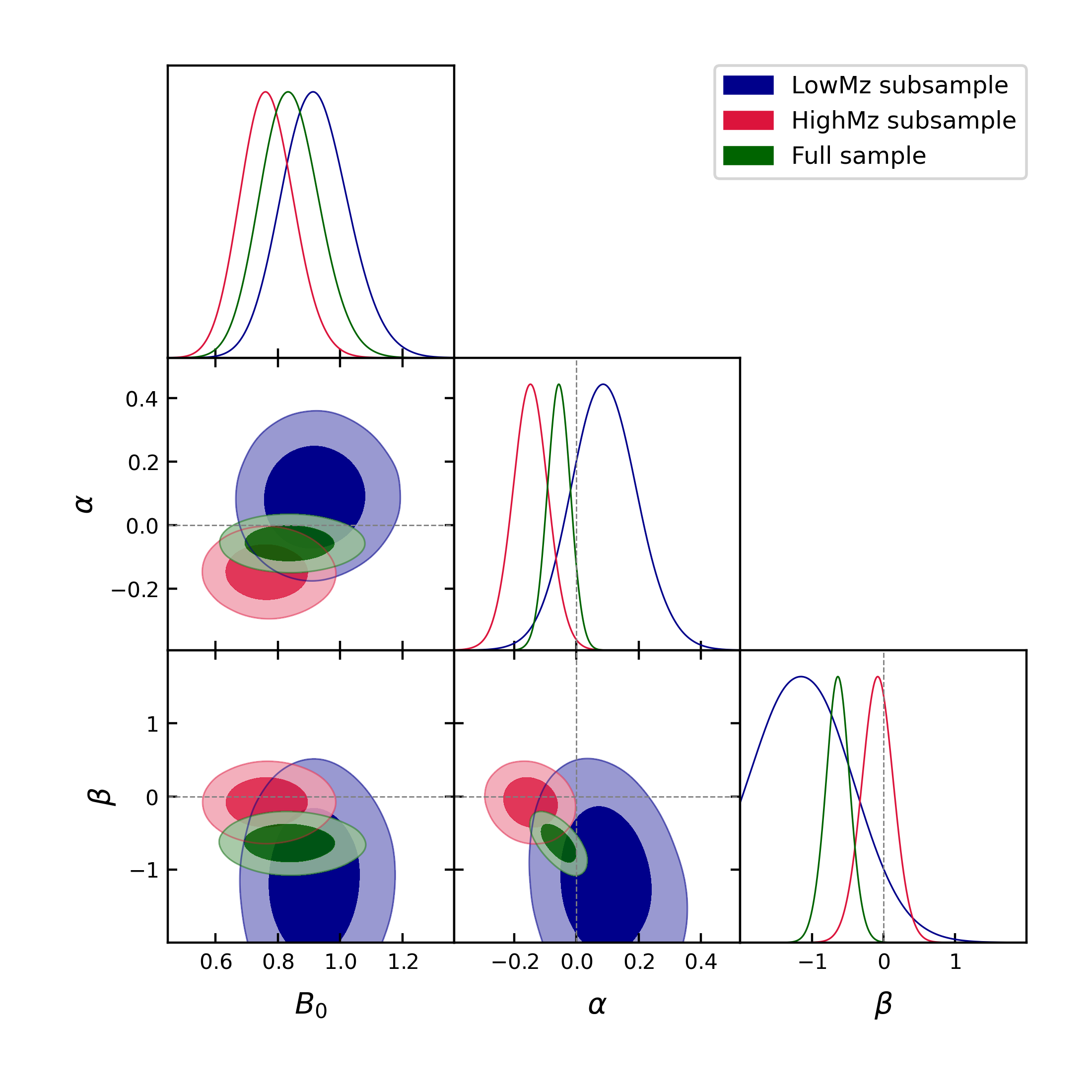}
    \caption{1D and 2D posterior distributions for the bias parameters in the three mass- and redshift-selected samples. The levels of the contours mark the 68\% and 95\% confidence levels. Gray dashed lines mark the reference values $(\alpha, \beta) = (0, 0)$.}
    \label{fig:contours_sample_dependence}
\end{figure}
\begin{figure*}
    \centering
    \includegraphics[width = \textwidth, trim = {0.5cm 2cm 0.2cm 0.25cm}, clip]{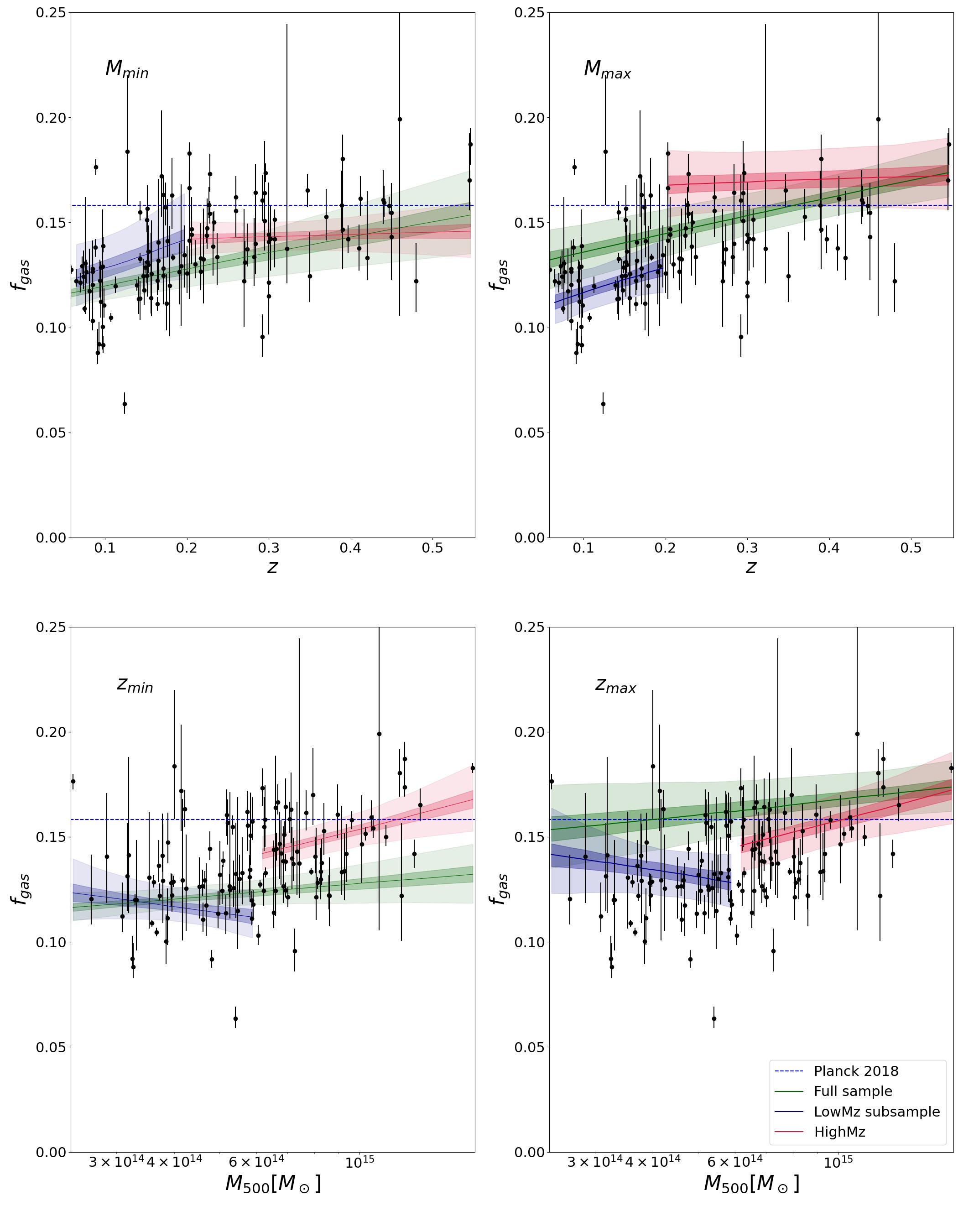}
    \caption{Fits obtained from our analysis, for the different mass- and redshift-selected samples. The results in the top two panels are represented with respect to redshift at constant mass (respectively minimal and maximal masses of the full sample), while the bottom two panels are the results presented with respect to mass, at a fixed redshift (respectively minimal and maximal redshifts of the full sample). The shaded areas around the curves mark the 68\% and 95\% confidence levels. The blue dashed line marks the reference value $\Omega_b/\Omega_m = 0.156$.}
    \label{fig:fits_sample_dependence}
\end{figure*}

If an evolution of the bias seems necessary to properly constrain cosmological parameters using $f_{gas}$ data at $R_{500}$, the strength of this evolution might differ depending on the masses and redshifts of the clusters we consider.
We thus repeat the previous study, this time focusing only on the bias parameters, using the previously defined {\it LowMz} and {\it HighMz} subsamples in addition to the full sample.
We recall that in this section, we are considering the VB + $\Omega_m$ scenario, with an additional prior on $\Omega_b/\Omega_m$.
The summary of our results is given in Table \ref{tab:results_sample_dependence} and in Figures \ref{fig:contours_sample_dependence} and \ref{fig:fits_sample_dependence}.

\begin{table*}[h!]
    \centering
    \caption{Constraints obtained on the bias parameters depending on the considered sample. Uncertainties are given at 68\% c.l.}
    \begin{tabular}{cccc}
    \hline \hline
    {\bf Parameter} & LowMz subsample & HighMz subsample & Full sample \\
    \hline
    $\mathbf{B_0}$ &  $0.92^{+0.10}_{-0.11}$ & $0.767\pm 0.086$ & $0.840\pm 0.095$ \\
    $\boldsymbol{\alpha}$ & $0.09\pm 0.11$ & $-0.149\pm 0.058$ & $-0.057\pm 0.038$ \\
    $\boldsymbol{\beta}$ & $-0.995^{+0.44}_{-0.77}$ & $-0.08\pm 0.23$ & $-0.64 \pm 0.18$ \\
    \hline
    \end{tabular}
    \label{tab:results_sample_dependence}
\end{table*}

Our results when considering the full sample are (as expected) the same as when adopting flat priors on the cosmological parameters. The only slight exception is $B_0$, peaking slightly higher due to the absence of a prior on the total bias. We find an amplitude of $ B_{0, full} = 0.840\pm 0.095$.
The parameters accounting for the bias evolution remain unchanged, with $\alpha_{full} = -0.057\pm 0.038$ and $\beta_{full} = -0.64 \pm 0.18$.
Thus, our results remain compatible with no mass evolution of the bias but still find a strong hint for a redshift dependence.

If all subsamples provide compatible values for the amplitude with $B_{0, lowMz} = 0.92^{+0.10}_{-0.11}$, $B_{0, HighMz} = 0.767\pm 0.086$ and $ B_{0, full} = 0.840\pm 0.095$, this cannot be said regarding $\alpha$ and $\beta$. 
Indeed, if the study of the full sample seems to suggest no mass evolution and a mild redshift trend of the bias, we observe the reverse behavior in the {\it HighMz} subsample. 
With $\alpha_{highMz} = -0.149\pm 0.058$ and $\beta_{HighMz} = -0.08\pm 0.23$, the preferred scenario would be the one of $B$ constant with redshift, yet decreasing with cluster mass.
On the other end of the mass-redshift plane, the results show exactly the opposite evolution, in agreement with the constraints from the full sample but aggravating the trends.
With $\alpha_{LowMz} = 0.09\pm0.11$ and $\beta_{LowMz} = -0.995^{+0.44}_{-0.78}$, we show we are fully compatible with no mass evolution of the bias, even with the posterior of $\alpha$ peaking slightly above 0 contrary to the other samples.
More importantly, we show a strong decreasing trend of $B$ with redshift, as we obtain $\beta$ peaking close to -1.
We note that the posterior distribution for this subsample is quite wider it is than for the full sample or the {\it HighMz} subsample. 
This is due to the smaller mass and redshift range of this sample (see Figure \ref{fig:binned_Mz}), diminishing its constraining power with respect to the other two selections.
This is, however, sufficient to highlight a redshift dependence of the bias when considering the least massive clusters of our sample at the lowest redshifts.

In Figure \ref{fig:fits_sample_dependence}, we show what these values of the bias parameters translate to in terms of the gas fraction with respect to redshift and mass.
We show these fits computed at $[M_{min}, M_{max}]$ with $z$ free and at $[z_{min}, z_{max}]$ with $M$ free, taking into account the uncertainties at 68\% and 95\% c.l.
We first notice what was highlighted in the contours of Figure \ref{fig:contours_sample_dependence}, which is that the results obtained for the full sample mainly fall in between the two extreme cases of the  {\it LowMz} and {\it HighMz} subsamples.
Secondly we show that the incompatibility between the two smaller subsamples is fully visible in the result of the fits.
In the bottom two panels showing the $f_{gas}(M)$ relation, the {\it LowMz} and {\it HighMz} even seem to exhibit opposite trends, similarly to what is seen in Figure \ref{fig:contours_sample_dependence}.
Finally, we note an offset in the relative positions of the curves for the two subsamples, depending on mass and redshift.
This is simply due to the fact that our model describes a simultaneous evolution of the bias both with mass and redshift, which happens to be non-zero in our case.

In summary, we claim to have found strong evidence for the sample dependence of our results regarding the mass and redshift evolution of the hydrostatic mass bias.
Such a sample dependence had already been noted in other works studying the evolution of the bias using tSZ cluster counts, but not when using $f_{gas}$ data.

\section{Discussion}
\label{sect:discussion}
As we have shown in the previous section, considering an evolution of the bias seems to be necessary to infer sensible cosmological constraints.
On the other hand, the variation of the bias that we measure is very dependent on the sample we consider, as we show different trends of the bias depending on mass and redshift selections inside our main sample.
We discuss these results here, trying to take into account all the systematic effects that could appear and bias our results.
We also compare our findings to previous studies focused on the bias and its evolution, from different probes.
\subsection{Possible sources of systematic effects}
\subsubsection{Instrumental calibration effects}
\label{sect:instrumental_calibration}
\begin{figure}
    \centering
    \includegraphics[width = 0.5\textwidth, trim = {0.25cm 0.75cm 1cm 2cm}, clip]{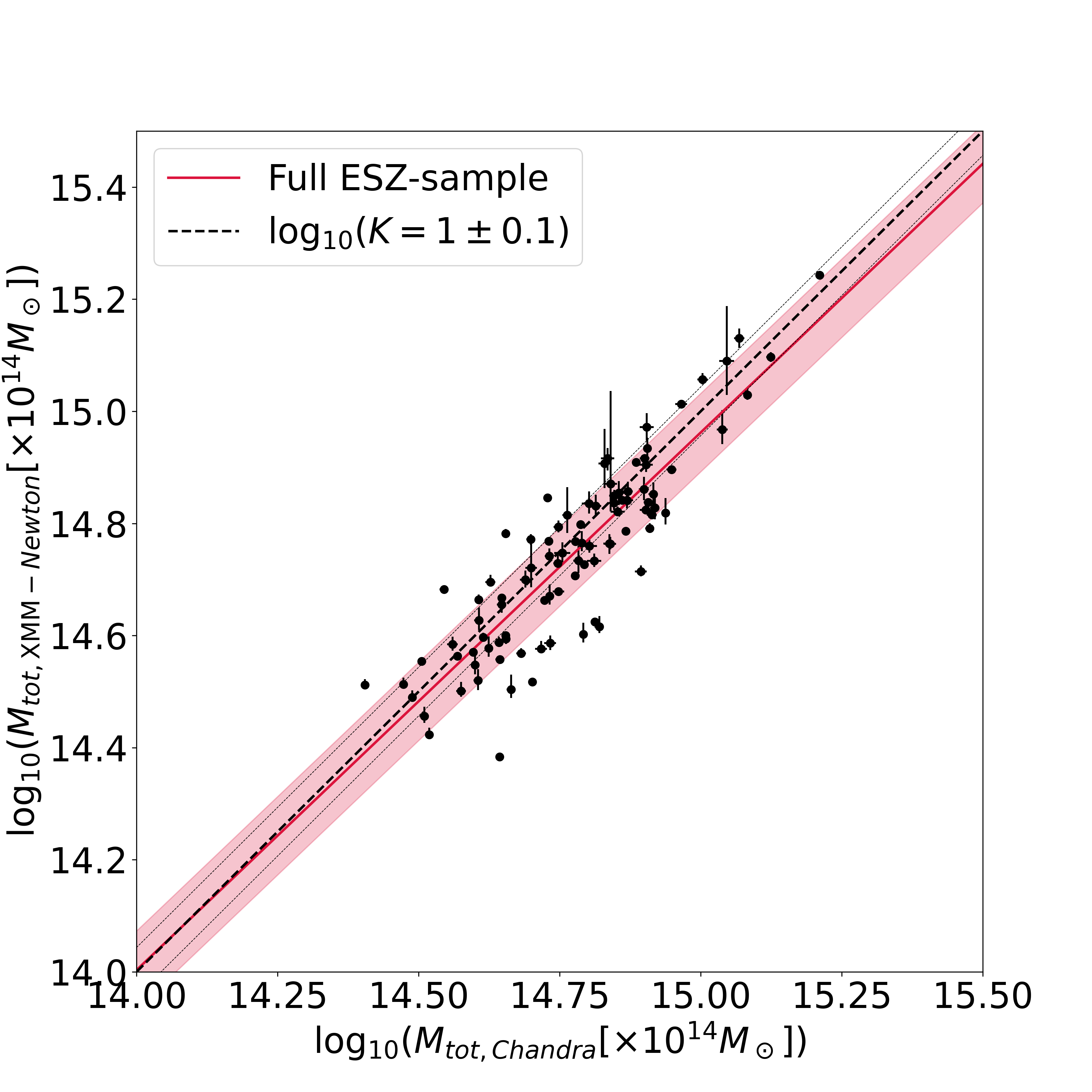}
    \caption{Comparison of the total masses inside the {\it Planck}-ESZ sample between XMM-{\it Newton} and {\it Chandra}.}
    \label{fig:XMM_v_Chandra}
\end{figure}

All of the masses used in this work were taken from \cite{2020ApJ...892..102L}, where the authors used XMM-{\it Newton} observations. 
Possible calibration effects may have impacted the mass measurements in their work and induced biases (see e.g. \cite{2013ApJ...767..116M} or \cite{2015A&A...575A..30S})
This effect has been taken into account under the form of a $K=1.0 \pm 0.1$ parameter in our analysis, but we try here to check if this assumption is sound.
The clusters of the {\it Planck}-ESZ sample have also been observed using {\it Chandra}, in \cite{2021ApJ...914...58A}.
From their work, we retrieved their total masses and compare them to the XMM-{\it Newton} masses. 
The result of the comparison is shown in Figure \ref{fig:XMM_v_Chandra}. 
We performed a fit of the point cloud in the log-log space, assuming a linear model of the form $\log(M_{XMM}) = a\log(M_{Chandra}) + b$.
Similarly to the model we defined in Equation \ref{eq:b_powerlaw} when studying the evolution of the bias, we set a pivot to the mean value of the {\it Chandra} masses, $\left <M_{Chandra} \right > = 6.32 \times 10^{14}M_\odot$.
We note that the results we present below do not change when putting the pivot at the median mass instead of the mean.
To perform the fit of the point cloud we resorted to an MCMC, in order to be able to account for the asymmetric measurement errors both in $X$ and $Y$ axis and the intrisic scatter of the data.
By doing this we obtain the following relation :
\begin{align}
    \log(M_{\mathrm{tot}, XMM}) = & (0.959 \pm 0.053) \log(M_{\mathrm{tot}, Chandra})\\
    \notag & - (0.029 \pm 0.008),
\end{align}
with an intrinsic scatter $\sigma_i = 15.2 \pm 1.3 \%$.

We show that we are fully compatible with a mass-independent mass calibration bias. 
We still however observe an offset, with the masses from XMM-{\it Newton} being globally lower than the {\it Chandra} masses. 
This result had already been shown previously (see Appendix D of \cite{2015A&A...575A..30S}).
This offset is however well accounted for by the 10\% uncertainty we allow in the $K$ prior in the majority of the sample.
We do, however, point out that this comparison has been carried out only on the total masses, while our study is focused on the gas mass fraction.
To be able to fully rule out possible biases from instrumental effects we would need to compare the gas mass fractions obtained from both observations rather than the total masses.
Unfortunately as we do not have access to the gas masses from the {\it Chandra} observations, we can only assume that the compatibility we see for $M_{tot}$ stays true for $f_{gas}$. 
This assumption is sensible, as we have
\begin{equation}
   \frac{f_{gas, Chandra}}{f_{gas, \mathrm{XMM}-Newton}} = \frac{M_{gas, Chandra}}{M_{gas, \mathrm{XMM}-Newton}} \times \frac{M_{tot, \mathrm{XMM}-Newton}}{M_{tot, Chandra}}.
\end{equation}
As it happens, the discrepancy between {\it Chandra} and XMM-{\it Newton} mass measurements originates mainly from the calibration of the temperature profiles \citep{2015A&A...575A..30S}.
As these profiles are not used in the computation of the gas masses and as the $R_{500}$ used in both studies are compatible, is it reasonable to assume $M_{gas, Chandra}/M_{gas, \mathrm{XMM}-Newton} \sim 1$.
We thus argue that the compatibility that we see between {\it Chandra} and XMM-{\it Newton} total mass measurements holds true for the gas fractions.
We therefore assume that if calibration biases are affecting our study, they have only minor effects on our results.

\subsubsection{Other contributions to the evolution of $f_{gas}$}
\label{sect:contributions_to_fgas}
In this work, we consider the evolution of the gas fraction as a probe for the evolution of the hydrostatic mass bias, as well as a cosmological probe. 
As discussed in Sect. \ref{sect:theoretical_model_data} and shown in Eq. \ref{eq:fgas_model} however, several physical and instrumental effects may vary with mass and redshift and play a role in the evolution of $f_{gas}$.

 The depletion factor $\Upsilon$ is fully degenerated with the hydrostatic mass bias, both regarding its value and its evolution with mass and redshift.
As such, we need a robust prior on this parameter to be able to disentangle its effects on the $f_{gas}$ measurements from those of the bias.
Following \cite{2013MNRAS.431.1487P}, here we considered a depletion at $R_{500}$ $\Upsilon = 0.85\pm 0.03$, with no evolution with mass nor redshift.
This value is lower than the value of $\Upsilon = 0.94 \pm 0.03$ from \cite{2019A&A...621A..40E}, however the use of different priors on the mean value of the depletion only affects the results on $B_0$, and does not change the results on the mass and redshift evolution of the bias, as we show in Appendix \ref{annex:C}.
Furthermore, the latter study shows no evolution of the depletion with mass or redshift.
\cite{2020MNRAS.498.2114H} found a depletion $\Upsilon \sim 0.8$ in the clusters of the {\sc Fable} simulations, this time again with no clear trend with mass and redshift for clusters with $M_{500} > 3\times 10^{14}M_\odot$, compatible with the prior assumed in this work.
We however stress once again that both parameters are heavily degenerated and that a shift in the value of $\Upsilon$ will produce a similar shift in the predicted value of $B_0$. 
As $B_0$ and $\Omega_b/\Omega_m$ are also degenerated, a shift in the value of $\Upsilon$ might also produce a shift in the predicted value of $\Omega_b/\Omega_m$.
However, this has no effect on our values of $\alpha$ and $\beta$, nor on our constraints on other cosmological parameters.

In addition, a deviation from self-similarity due to baryonic physics may cause the $M_{tot}-M_{gas}$ relation to be mass- and redshift-dependent.
In \cite{2020ApJ...892..102L}, deviations from self-similarity of the ESZ cluster sample are studied.
In particular, the redshift evolution of the relation is compatible with 0, in agreement with the self-similar prediction.
On the other hand, the slope of the relation (i.e. its mass evolution), which we call $\gamma$ departs from the self similar prediction by more than $4\sigma$, with $\gamma = 0.802\pm 0.049$.
This result however was obtained on the {\it hydrostatic} masses computed from the X-ray observations and, thus, the fitted relation is $M_{HE}-M_{gas}$.
We show in Appendix \ref{annex:B} that the observed evolution of $f_{gas}$ with mass is, consequently, most likely due to a combination of an evolution of the bias and a deviation of the true $M_{tot}-M_{gas}$ relation from self-similarity.
However, we cannot disentangle the two effects and acknowledge that assuming or not self-similarity may lead to changes in the value of the mass evolution of the bias (see e.g. \cite{2016A&A...592A..12E} or \cite{2018MNRAS.474.4089T}).

The calibration bias, $K$, may evolve with cluster mass. 
We show in Sect. \ref{sect:instrumental_calibration} that when comparing {\it Chandra} and XMM-{\it Newton} masses such an evolution should be small. 
For this reason, any possible impact is accounted for by our prior on $K$. 

The stellar fraction may also vary both with mass and with redshift (see e.g. \cite{2012ApJ...745L...3L} or \cite{2020MNRAS.498.2114H}).
The mean value of this parameter being negligible, such an evolution only has a minor impact on our our results and we can consider it constant in our analysis, following the prior from \cite{2019A&A...621A..40E}.

Thus, the only two possible remaining contributions to the evolution of $f_{gas}$ are the hydrostatic bias, which can evolve with mass and redshift, and the cosmology, inducing a redshift evolution on the part of the gas fraction. 
Following our motivated choice of priors on both bias and cosmological parameters in the two parts of the analysis, we conclude that the only effect able to induce a mass and redshift evolution of $f_{gas}$ is the hydrostatic mass bias.
\subsubsection{Role of the parametrization}
\label{sect:parametrization}
\begin{figure}
    \centering
    \includegraphics[width = 0.5\textwidth, trim = {0.75cm 0.75cm 0.75cm 0.75cm}, clip]{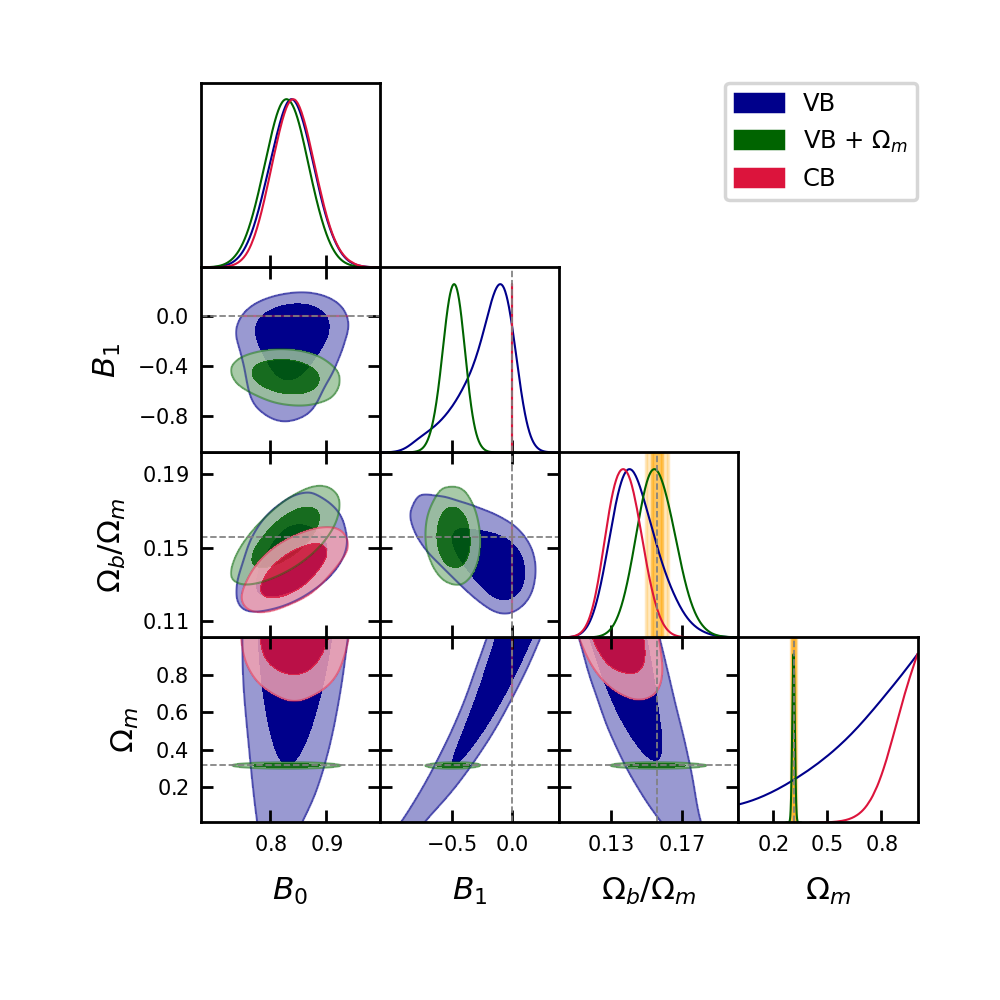}
    \caption{1D and 2D posteriors obtained when comparing the CB, VB and VB + $\Omega_m$ scenarios with a linear parametrization. The contours mark the 68\% and 95\% c.l. and the gray dashed lines highlight reference values for $(B_1, \Omega_b/\Omega_m, \Omega_m) = (0, 0.156, 0.315)$. The orange bands mark the \cite{2020A&A...641A...6P} values for $\Omega_b/\Omega_m$ and $\Omega_m$ at 2$\sigma$ c.l.}
    \label{fig:constant_v_varying_linear}
\end{figure}
\begin{figure}
    \centering
    \includegraphics[width = 0.5\textwidth, trim = {0.75cm 0.75cm 0.5cm 0.75cm}, clip]{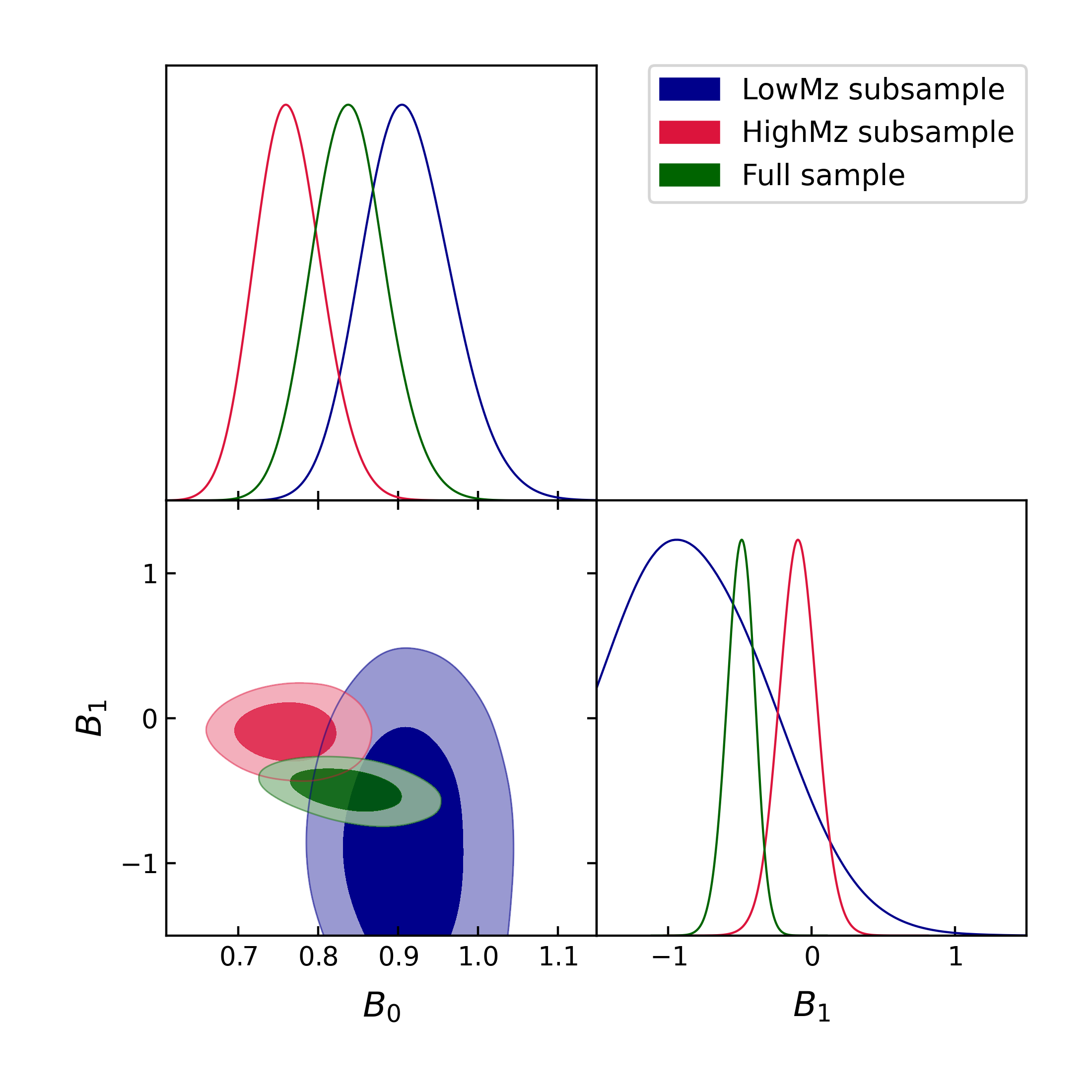}
    \caption{1D and 2D posteriors obtained when comparing the bias parameters derived for the three samples, when considering a linear evolution of the bias. The contours mark the 68\% and 95\% c.l.}
    \label{fig:sample_dependence_linear}
\end{figure}

\begin{table*} [h!]
    \centering
    \caption{Constraints obtained on bias and cosmological parameters, when assuming a linear parametrization. Uncertainties are given at 68\%c.l.}
    \begin{tabular}{ccccc}
    \hline \hline
     & $B_0$ & $B_1$ & $\Omega_b/\Omega_m$ & $\Omega_m$ \\
    \hline
    {\bf CB} & $0.842 \pm 0.038$ & 0 & $ 0.1375\pm0.0096$ & $ > 0.869$ \\
    {\bf VB} & $0.838\pm 0.040$ & $-0.21^{+0.27}_{-0.13}$ & $0.144^{+0.011}_{-0.015}$ & -- \\
    {\bf VB + $\Omega_m$} & $0.829\pm 0.039$ & $ -0.486\pm 0.093$ & $ 0.15\pm 0.011$ & $0.315\pm 0.007$ {\it (prior)}\\
    \hline
    {\bf LowMz subsample} & $0.912^{+0.051}_{-0.059}$ & $ -0.73^{+0.30}_{-0.66}$ & $0.156\pm 0.003$ {\it (prior)}& $0.315\pm 0.007$ {\it (prior)}\\
    {\bf HighMz subsample} & $0.763_{-0.044}^{+0.039}$ & $-0.095\pm0.14$ & $0.156\pm 0.003$ {\it (prior)}& $0.315\pm 0.007$ {\it (prior)}\\
    {\bf Full sample} & $0.839\pm 0.046$ & $-0.496^{+0.10}_{-0.091}$ & $0.156\pm 0.003$ {\it (prior)}& $0.315\pm 0.007$ {\it (prior)}\\
    \hline \hline
    \end{tabular}
    \tablefoot{We remind that when investigating the sample dependence (second part of the table), we are considering the VB + $\Omega_m$ scenario with a prior on $\Omega_b/\Omega_m$.}
    \label{tab:results_linear}
\end{table*}

In the previous work \cite{2022EPJWC.25700046W}, we investigated the evolution of $B$ with redshift, using a linear parametrization for the evolution of the bias,
\begin{equation}
    B(z) = B_0 + B_1(z-\left < z \right>)
\end{equation}
instead of a power law.
Here, we compare the results given by this choice of parametrization to the results we got for a power law model.
In this comparison of the linear case with the power law case we focus on the redshift dependence, as we did not find strong evidence for a mass evolution of the bias in the majority of our samples.
We show in Figures \ref{fig:constant_v_varying_linear} and \ref{fig:sample_dependence_linear} and in Table \ref{tab:results_linear} that the results are qualitatively consistent between the power law and linear descriptions.
Indeed, in both cases, when considering the VB + $\Omega_m$ scenario, we observe a strong sample dependence of the results.
As a matter of fact, we show in Figure \ref{fig:constant_v_varying_linear} that the {\it LowMz} clusters strongly favor a non-zero slope, namely, $B_1 = -0.73^{+0.30}_{-0.66}$.
On the high end of the mass-redshift plane, we find results that are consistent with the power law case and that are also compatible with no redshift evolution of $B$, as we find $B_1 = -0.095\pm 0.14$.
Our estimates of $B_0$ are also consistent between the power law case and the linear case for each subsample respectively, as we find $B_{0, lowMz} = 0.912_{-0.059}^{+0.051}$, $B_{0, highMz} = 0.763_{-0.039}^{+0.044}$ and $B_{0, full} = 0.839\pm 0.046$.

When studying the VB scenario, we still observe the strong degeneracy between the term accounting for the redshift evolution of $B$ (here, $B_1$) and $\Omega_m$.
This explains why we again obtained aberrant values of $\Omega_m$ when considering the CB scenario.
This compatibility between the qualitative results given by both parametrizations would hint at the fact that our results are not dominated by our choice of model for the bias evolution.



\subsection{Comparison with other studies}
\subsubsection{Mass and redshift trends of the bias}

\begin{figure}
    \centering
    \includegraphics[width= 0.5\textwidth, trim={0.25cm 0.5cm 1.25cm 3.5cm}, clip]{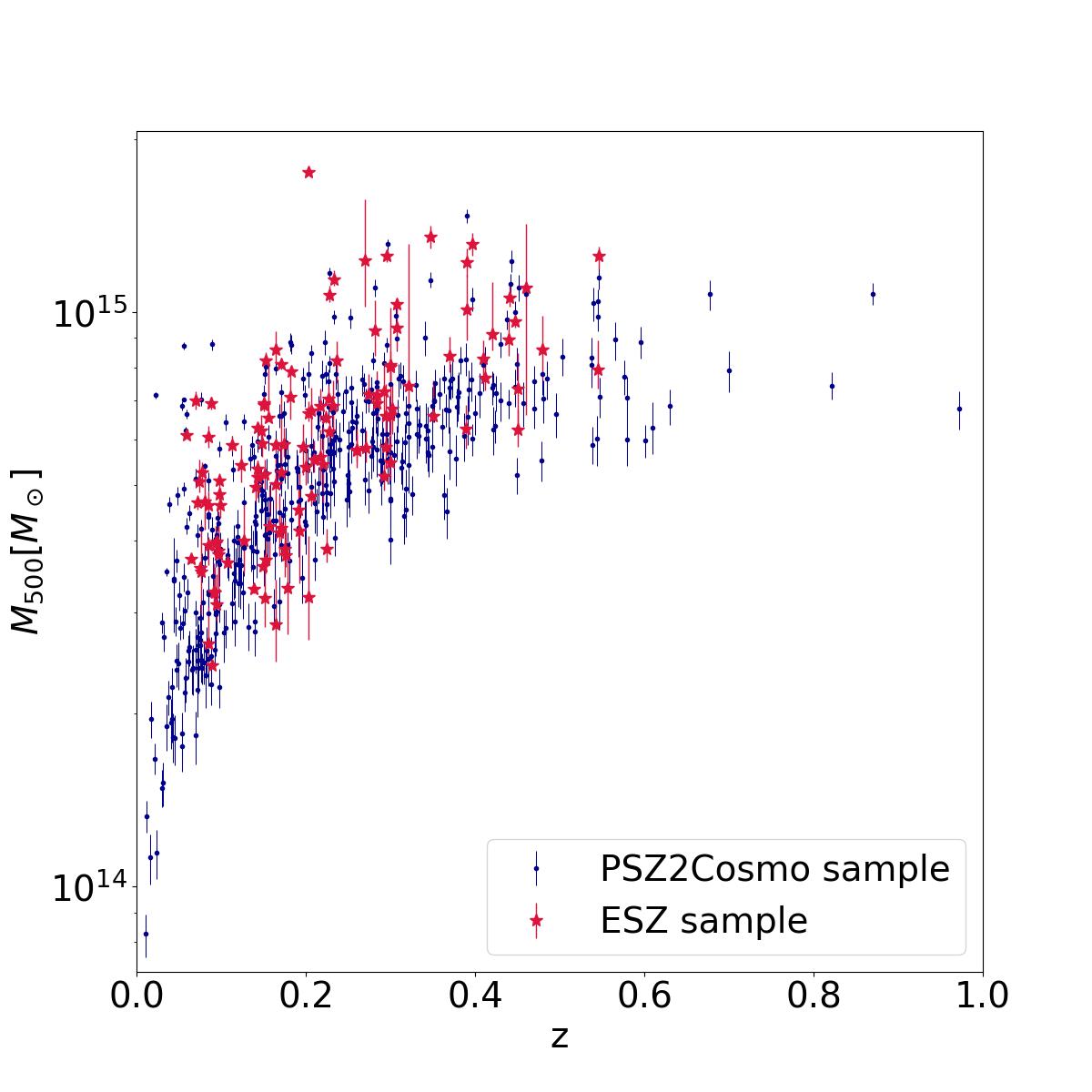}
    \caption{Comparison between the PSZ2Cosmo sample used in \cite{2019A&A...626A..27S} and the {\it Planck}-ESZ sample.}
    \label{fig:PSZ2C_v_ESZ}
\end{figure}

In this work, we show a strong sample dependence of our sampled results on different mass and redshift variations of the bias depending on the mass and redshift range of the subsample.
We show that for low redshift and low mass clusters, the bias tends to stay constant with mass, while it increases toward higher redshift. 
The reverse behavior is observed in the case of high masses and high redshift clusters, with a bias that is constant with redshift but slightly increasing with the mass of the objects.
This sample dependence had already been noted in \cite{2019A&A...626A..27S}, where the authors used tSZ number counts on the cosmology sample of the PSZ2 catalog. 
The authors indeed noted a non-zero trend of the bias with redshift when considering their complete sample, which disappeared when considering only the clusters above $z = 0.2$.
We note, however, that the compatibility of our results with that study regarding the mass and redshift trends depends on the considered subsample of clusters.
We observe the same behavior regarding the value of the constant term of the bias $B_0$, with compatible results for the high redshift clusters but incompatible ones when considering the full sample.
We might argue that these differences come from different choices of priors in the two studies.
Indeed, when investigating sample dependencies we did not assume any prior on the bias parameters and we considered priors on cosmology, whereas \cite{2019A&A...626A..27S} assumed a prior on the total value of the bias and let their cosmological parameters remain free.
However, this might not have such a strong impact, since in the first part in the analysis, we let the cosmological parameters free and we put a prior on the total value of the bias; in our results we do not see significant deviations in the value of $B_0$ between the two cases.
These differences are actually most probably due to differences between the clusters considered in the two studies, as the clusters from the ESZ sample have globally higher mass at an equivalent redshift than the PSZ2Cosmo sample considered in their study, as we show in Figure \ref{fig:PSZ2C_v_ESZ}.
If anything, this could be yet another indication of the strong sample dependence in the results we present here.

Our results are nonetheless consistent with other weak lensing studies, which include Weighing the Giants (WtG, \cite{2014MNRAS.443.1973V} or CCCP \citep{2015MNRAS.449..685H}, displaying a mild decreasing trend in mass for $B$ when considering high redshift ($z_{WtG} = 0.31$, $z_{CCCP} = 0.246$) and high mass ($M_{500, WtG} = 13.58\times10^{14}M_\odot$, $M_{500, CCCP} =10.38\times10^{14}M_\odot$) clusters.
More recent works such as the X-ray study X-COP \citep{2019A&A...621A..40E} also seem to show a possible mass-dependent bias, with a decreasing trend of $B$. 
The weak-lensing studies {\sc LoCuSS} \citep{2016MNRAS.456L..74S} and {\sc CoMaLit} \citep{2017MNRAS.468.3322S} both find decreasing trends of $B$ with redshift, in agreement with this work, for clusters in the mass and redshift range of our sample ($z_{{\sc LoCuSS}} = 0.22$, $M_{{\sc LoCuSS}} = 6.8\times10^{14}M_\odot$).

On a side note, our results regarding the amplitude, $B_0$, are also compatible with \cite{2020A&A...644A..78L}, where the authors measured the ratio of hydrostatic to weak lensing mass $M_{HE}/M_{WL}$ inside the {\it Planck}-ESZ sample. 
The authors found a ratio $(M_{HE}/M_{WL})_{ESZ} = 0.74 \pm 0.06$, which is in agreement with our amplitude $B_{0, ESZ} = 0.840 \pm 0.095$.

\subsubsection{Discussing the choice of a sample at $R_{500}$}
\label{sect:comparison_R2500}
In this work, we focus on gas fractions taken at $R_{500}$, which is larger than most works using $f_{gas}$ as a cosmological probe, carried out at $R_{2500}$. 
The choice of $R_{2500}$ is generally motivated by the low scatter of the gas fraction data around those radii (see \cite{2014MNRAS.440.2077M} and references therein), allowing for more precise cosmological constraints.
The scatter $f_{gas}$ in data is larger at $R_{500}$, however, this inconvenient is balanced by the stability of the depletion factor $\Upsilon$ at this radius.
Indeed, as shown by the hydrodynamical simulations from \cite{2013MNRAS.431.1487P}, the value of $\Upsilon$ does not vary much with the radius when it is measured around $R_{500}$.
This reduces the possibility of a biased estimation of the depletion due to incorrect determinations of $R_{500}$.
On the other hand $\Upsilon$ starts to decrease in the vicinity of $R_{2500}$.
As a consequence, if the radius is not properly measured, (e.g. due to erroneous estimations of the density contrast), the estimation of the depletion will be biased.
Our purpose in this study being to constrain the hydrostatic mass bias and its evolution, we need a robust prior on the depletion factor, due to the degeneracy between $\Upsilon$ and $B$.
A bias in the value of the depletion due to an incorrectly measured radius would then impact our results on the mean value and evolution of $B$.

\subsection{Discussing the tension on the bias value}
\begin{figure}
    \centering
    \includegraphics[width= 0.475\textwidth, trim={0.48cm 0cm 0cm 0.35cm}, clip]{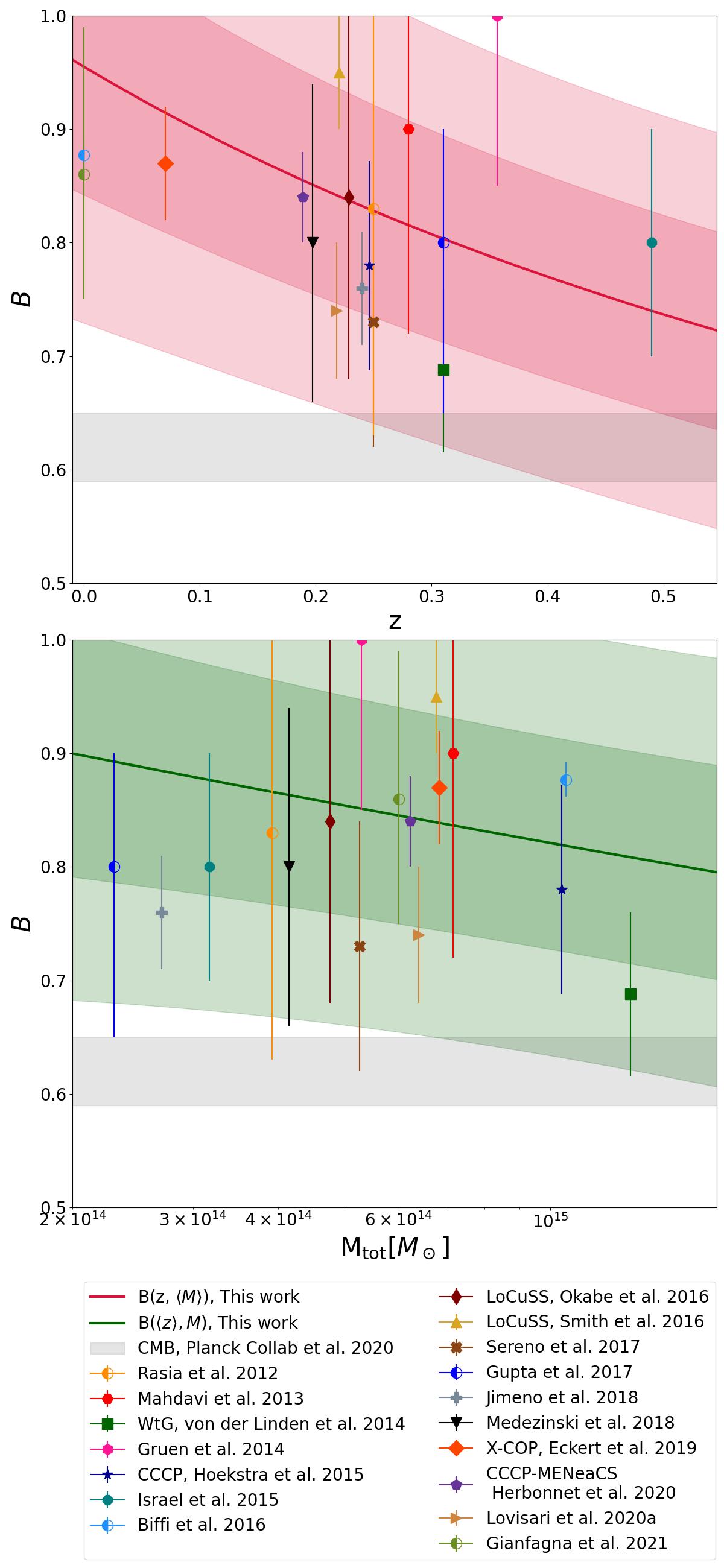}
    \caption{Comparison of our value of the mass bias depending on mass and redshift with other works in X-ray, weak lensing, and hydrodynamical simulations.
    In both panels, the shaded areas mark the 1 and 2$\sigma$ confidence levels and
    the gray band marks the value preferred by CMB observations of $B = 0.62 \pm 0.003$. 
    The half-filled circles are hydrodynamical simulation works. 
    Other points represent works based on observations.
    {\it Top }: Value of the bias depending on redshift, computed for the mean mass of our sample. 
    {\it Bottom }: Value of the bias depending on mass, computed at the mean redshift of our sample.\\
    {\bf References :} \cite{2012NJPh...14e5018R}, \cite{2013ApJ...767..116M}, \cite{2014MNRAS.443.1973V}, \cite{2014MNRAS.442.1507G}, \cite{2015MNRAS.449..685H}, \cite{2015MNRAS.448..814I}, \cite{2016ApJ...827..112B}, \cite{2016MNRAS.461.3794O}, \cite{2016MNRAS.456L..74S}, \cite{2017MNRAS.472.1946S}, \cite{2017MNRAS.469.3069G}, \cite{2018MNRAS.478..638J}, \cite{2018PASJ...70S..28M}, \cite{2019A&A...621A..40E}, \cite{2019A&A...626A..27S}, \cite{2020MNRAS.497.4684H}, \cite{2020A&A...644A..78L}, \cite{2020A&A...641A...6P}, \cite{2021MNRAS.502.5115G}.}
    \label{fig:B(z, M)}
\end{figure}
\begin{figure}
    \centering
    \includegraphics[width = 0.5\textwidth, trim = {0.75cm 0.65cm 2cm 1.75cm}, clip]{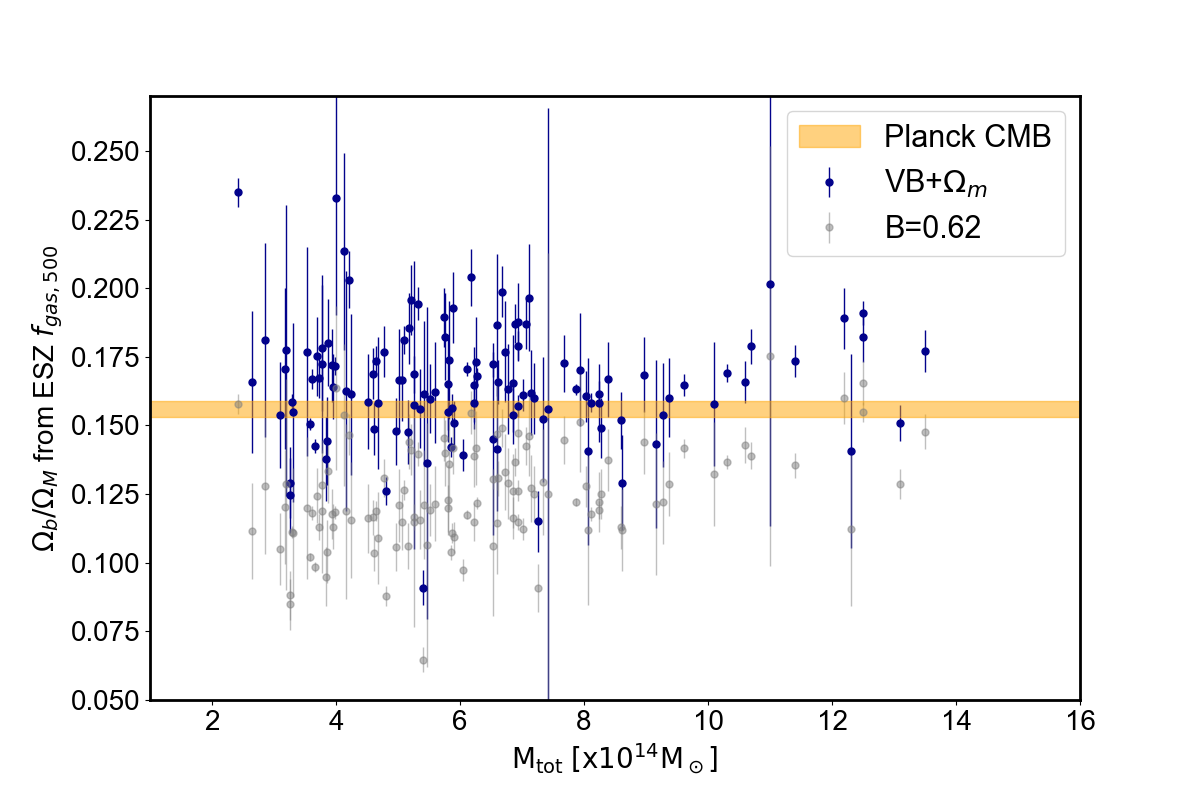}
    \caption{Comparison of the expected baryon fraction $\Omega_b/\Omega_m$ between the bias we derived in the VB + $\Omega_m$ scenario and $B = 0.62$ from \citep{2020A&A...641A...6P}.}
    \label{fig:ObOm_predictions_B0.6}
\end{figure}

Finally, we highlight that the value of {\bf $B_0 = 0.840 \pm 0.095$} we found for the full sample is in agreement with a collection of other studies, including the aforementioned weak-lensing and X-ray studies, as well as with works from hydrodynamical simulations, as shows in Figure \ref{fig:B(z, M)}. 
The works shown in Figure \ref{fig:B(z, M)} are those for which the bias is known at $R_{500}$ with a central value and error bars, where the mean mass and redshift of the samples are available.
We show, however, that this value is incompatible with the values $B_0 = 0.62 \pm 0.03$ from \cite{2020A&A...641A...6P}, or $B_0 \lesssim 0.67$ from \cite{2018A&A...614A..13S}, which are needed to reconcile local measurements of the bias with the combination of CMB and tSZ number counts measurements. 
Indeed, as we show in Figure \ref{fig:B(z, M)}, the tension is alleviated only for the highest redshifts and, more particularly for the highest masses, for clusters with $M \gtrsim 10^{15}M_\odot$.

Similarly the X-COP study \citep{2019A&A...621A..40E} shows that assuming a bias of $B = 0.58 \pm 0.04$ from \cite{2016A&A...594A..24P} results in gas fractions that are way lower than expected, as they find a median gas fraction of their sample $f_{gas, B=0.58} = 0.108\pm 0.006$. 
This would imply that the galaxy clusters from their sample are missing about a third of their gas.
A low value for the bias thus seems highly unlikely given its implications fir the gas content of galaxy clusters.

We show a similar result in this work.
Indeed, from Equation \ref{eq:fgas_model}, it is possible to compute the expected universal baryon fraction for a certain value of the bias, namely :
\begin{equation}
    \label{eq:Ob_Om_wrt_B}
    \frac{\Omega_b}{\Omega_m} = \left(f_{gas} + f_* \right) \frac{1}{KA(z)}\frac{B(M,z)}{\Upsilon(M,z)}\left( \frac{D_A(z)}{D_A(z)^{ref}}\right)^{3/2},
\end{equation}
meaning $\Omega_b/\Omega_m \propto B$.
When assuming a low value of the bias $B = 0.62 \pm 0.03$ from \cite{2020A&A...641A...6P}, we show that the derived baryon fraction becomes $\Omega_b/\Omega_m = 0.108 \pm 0.018$, which is incompatible with the value of the baryon fraction derived from the CMB, $\Omega_b/\Omega_m = 0.156 \pm 0.003$ \citep{2020A&A...641A...6P}, as we show in Figure \ref{fig:ObOm_predictions_B0.6}.
Assuming a low value of the bias would then imply that the Universe hosts roughly 20\% less baryons than expected.
We thus argue that a value of the bias $B = 0.62 \pm 0.03$ seems highly unlikely provided our gas fraction data.

\section{Conclusion}
\label{sect:conclusions}
We measured the gas mass fraction of galaxy clusters at $R_{500}$ and used it to constrain a possible mass and redshift evolution of the hydrostatic mass bias.
To do so, we compared the cosmological constraints we obtained when assuming a varying bias to those we obtained when assuming a constant $B$.
We show that assuming a redshift evolution seems necessary when performing a cosmological analysis using $f_{gas}$ data at $R_{500}$.
Indeed we show a significant degeneracy between the redshift dependence of the bias $\beta$ and the matter density $\Omega_m$.
This degeneracy implies that high and close to zero $\beta$ push the matter density to higher values. 
As a consequence, assuming a constant bias implies $\Omega_m > 0.860$, which is aberrant.
Forcing $\Omega_m$ to have sensible values by imposing a \cite{2020A&A...641A...6P} prior in turns induces $\beta = -0.64 \pm 0.17$, in a $3.8 \sigma$ tension with 0.

We show, however, that these results are strongly dependent on the considered sample.
Indeed for the least massive clusters of our sample at the lowest redshifts, we show an important decreasing trend of $B$ with redshift and no trend with mass, with the set $(\alpha, \beta) = (0.09 \pm 0.11, -0.995_{-0.78}^{+0.44})$.
On the other hand, for the most massive clusters at highest redshifts, we observe no variation with redshift but we see that a decreasing evolution with mass is favored, with $(\alpha, \beta) = (-0.149 \pm 0.058, -0.08 \pm 0.23)$.
When we consider the full sample, the results we obtain lie between the two extremes, largely favoring a decreasing trend of $B$ with redshift, yet remaining compatible with no mass trend of the bias as we obtain  $(\alpha, \beta) = (-0.057 \pm 0.038, -0.64 \pm 0.18)$. 
We recall, however, that other selection effects might affect our results as the clusters at the highest redshifts are also generally the most massive.
In addition, deviations from self-similarity may lead to changes in the value of the mass evolution of the bias.

In order to identify and rule out different sources of systematic effects in our study, we looked at the possibility of being subject to instrumental calibration effects. 
Using mass measurements of the galaxy clusters in our sample both from {\it Chandra} and XMM-{\it Newton}, we find no evidence of a bias that would significantly change our results.
In this study, we assumed a power-law description for the evolution of the bias.
We look at the effect of this choice of parametrization by comparing our results to those we obtain when assuming a linear evolution of $B$ with redshift. 
We find no major difference with the power law case, as we still observe a strong degeneracy between the redshift evolution of $B$ and $\Omega_m$, favoring an evolution of the bias with redshift when considering a {\it Planck} prior on $\Omega_m$.
Furthermore, the sample dependence we observe for the power law case holds true when assuming a linear evolution of the bias.

Despite these results, we stress that our results remain compatible with a collection of X-ray, weak lensing, and hydrodynamical simulation works regarding the mean value of the bias, given our finding of $B_0 = 0.840 \pm 0.095$. 
This value remains, on the other hand, in tension with the value required from the combination of CMB observations and tSZ cluster counts to alleviate the tension on $\sigma_8$.

Finally, we recall that this work is focused on gas fractions taken at $R_{500}$, with a goal to obtain constraints on two cosmological parameters, the universal baryon fraction, $\Omega_b/\Omega_m$, and the matter density of the Universe, $\Omega_m$, which are the two parameters mainly probed by $f_{gas}$.
We thus argue that the gas fraction can be used to set constraints on the cosmological model, albeit provided that one properly models the gas effects taking place inside clusters and provided that $f_{gas}$ is used in combination with other probes.
 
\begin{acknowledgements}
      The authors thank the anonymous referee for their helpful comments and discussion.
      They also acknowledge the fruitful discussions and comments from Lorenzo Lovisari, Stefano Ettori, Hideki Tanimura, Daniela Galárraga-Espinosa, Edouard Lecoq, Joseph Kuruvilla and Celine Gouin.
      The authors also thank the organisers and participants of the 2021 edition of the "Observing the mm Universe with the NIKA2 Camera" conference for their useful comments and discussions. RW acknowledges financial support from the Ecole Doctorale d'Astronomie et d'Astrophysique d'Ile-de-France (ED AAIF).
      This research has made use of the SZ-Cluster Database operated by the Integrated Data and Operation Center (IDOC) at the Institut d'Astrophysique Spatiale (IAS) under contract with CNES and CNRS.
      This project was carried out using the Python libraries {\tt matplotlib} \citep{2007CSE.....9...90H}, {\tt numpy} \citep{harris2020array}, {\tt astropy} (\cite{astropy:2013}, \cite{astropy:2018}) and {\tt pandas} \citep{jeff_reback_2021_4681666}. It also made use of the Python library for MCMC sampling {\tt emcee} \citep{2013PASP..125..306F}, and of the {\tt getdist} \citep{2019arXiv191013970L} package to read posterior distributions.
\end{acknowledgements}

%
%

\bibliographystyle{aa} 
\bibliography{fgas_bias} 

\begin{appendix}
\section{Cluster sample}
\label{annex:A}
The redshifts, gas masses, total masses, and gas mass fractions within $R_{500}$ of the clusters we used for this study are given in Table \ref{tab:cluster_sample}. The redshifts and gas and total masses for the clusters were taken from \cite{2020ApJ...892..102L}. We computed the gas fractions from these masses. 

\longtab[1]{
\begin{longtable}{c c c c c}
\caption{The 120 clusters we used in this study, from the {\it Planck}-ESZ sample. }\\
\hline \hline 
Planck Name & z & $M_{gas, 500} [\times 10^{14}M_\odot]$ & $M_{tot, 500} [\times 10^{14}M_\odot]$ &  $f_{gas, 500}$ \\ 
\hline 
\endfirsthead
\multicolumn{5}{c}{Table \ref{tab:cluster_sample} continued from previous page} \\ 
\hline \hline 
Planck Name & z & $M_{gas, 500} [\times 10^{14}M_\odot]$ & $M_{tot, 500} [\times 10^{14}M_\odot]$ &  $f_{gas, 500}$ \\ 
\hline
\endhead
\hline
\endfoot
\hline
\endlastfoot
G000.44-41.83 &  0.165 &        $0.661_{-0.033}^{+0.036}$ &           $5.01_{-0.48}^{+0.55}$ & $0.132_{-0.014}^{+0.016}$ \\
G002.74-56.18 &  0.141 &        $0.563_{-0.014}^{+0.021}$ &           $4.96_{-0.28}^{+0.43}$ & $0.114_{-0.007}^{+0.011}$ \\
G003.90-59.41 &  0.151 &        $0.865_{-0.008}^{+0.008}$ &           $6.94_{-0.19}^{+0.19}$ & $0.125_{-0.004}^{+0.004}$ \\
G006.70-35.54 &  0.089 &        $0.427_{-0.008}^{+0.005}$ &  $  2.42_{-0.03}^{+0.04}$ & $0.176_{-0.004}^{+0.004}$ \\
G006.78+30.46 &  0.203 &        $3.200_{-0.020}^{+0.020}$ &          $17.50_{-0.20}^{+0.20}$ & $0.183_{-0.002}^{+0.002}$ \\*
G008.44-56.35 &  0.149 &        $0.464_{-0.004}^{+0.005}$ &           $3.61_{-0.07}^{+0.08}$ & $0.129_{-0.003}^{+0.003}$ \\
G008.93-81.23 &  0.307 &        $1.560_{-0.010}^{+0.010}$ &          $10.30_{-0.20}^{+0.20}$ & $0.151_{-0.003}^{+0.003}$ \\
G021.09+33.25 &  0.151 &        $1.040_{-0.030}^{+0.030}$ &           $6.88_{-0.18}^{+0.20}$ & $0.151_{-0.006}^{+0.006}$ \\
G036.72+14.92  &  0.152 &        $0.816_{-0.028}^{+0.029}$ &           $5.21_{-0.29}^{+0.32}$ & $0.157_{-0.010}^{+0.011}$ \\
G039.85-39.98 &  0.176 &        $0.532_{-0.017}^{+0.036}$ &           $3.77_{-0.17}^{+0.47}$ & $0.141_{-0.008}^{+0.020}$ \\
G042.82+56.61 &  0.072 &        $0.601_{-0.007}^{+0.008}$ &           $4.65_{-0.13}^{+0.15}$ & $0.129_{-0.004}^{+0.005}$ \\
G046.08+27.18 &  0.389 &        $0.990_{-0.030}^{+0.035}$ &           $6.26_{-0.48}^{+0.61}$ & $0.158_{-0.013}^{+0.016}$ \\
G046.50-49.43 &  0.085 &        $0.624_{-0.008}^{+0.009}$ &           $6.05_{-0.26}^{+0.28}$ & $0.103_{-0.005}^{+0.005}$ \\
G049.20+30.86 &  0.164 &        $0.651_{-0.007}^{+0.007}$ &           $5.86_{-0.16}^{+0.16}$ & $0.111_{-0.003}^{+0.003}$ \\
G049.33+44.38 &  0.097 &        $0.385_{-0.017}^{+0.018}$ &           $3.84_{-0.38}^{+0.46}$ & $0.100_{-0.011}^{+0.013}$ \\
G049.66-49.50 &  0.098 &        $0.441_{-0.008}^{+0.008}$ &           $4.81_{-0.19}^{+0.21}$ & $0.092_{-0.004}^{+0.004}$ \\
G053.52+59.54 &  0.113 &        $0.703_{-0.008}^{+0.010}$ &           $5.87_{-0.15}^{+0.20}$ & $0.120_{-0.003}^{+0.004}$ \\
G055.60+31.86 &  0.224 &        $0.941_{-0.010}^{+0.012}$ &           $6.54_{-0.15}^{+0.18}$ & $0.144_{-0.004}^{+0.004}$ \\
G055.97-34.88 &  0.124 &        $0.344_{-0.009}^{+0.009}$ &           $5.41_{-0.37}^{+0.45}$ & $0.064_{-0.005}^{+0.006}$ \\
G056.81+36.31 &  0.095 &        $0.512_{-0.004}^{+0.004}$ &           $3.98_{-0.07}^{+0.08}$ & $0.129_{-0.002}^{+0.003}$ \\
G056.96-55.07 &  0.447 &        $1.520_{-0.010}^{+0.010}$ &           $9.62_{-0.25}^{+0.25}$ & $0.158_{-0.004}^{+0.004}$ \\
G057.26-45.35 &  0.397 &        $1.860_{-0.020}^{+0.020}$ &          $13.10_{-0.60}^{+0.60}$ & $0.142_{-0.007}^{+0.007}$ \\
G058.28+18.59 &  0.065 &        $0.454_{-0.005}^{+0.020}$ &           $3.72_{-0.07}^{+0.06}$ & $0.122_{-0.003}^{+0.006}$ \\
G062.42-46.41 &  0.091 &        $0.287_{-0.012}^{+0.013}$ &           $3.26_{-0.15}^{+0.39}$ & $0.088_{-0.005}^{+0.011}$ \\
G067.23+67.46 &  0.171 &        $1.040_{-0.010}^{+0.000}$ &           $8.11_{-0.18}^{+0.17}$ & $0.128_{-0.003}^{+0.003}$ \\
G071.61+29.79 &  0.157 &        $0.532_{-0.043}^{+0.051}$ &           $4.24_{-0.59}^{+0.76}$ & $0.125_{-0.020}^{+0.026}$ \\
G072.63+41.46 &  0.228 &        $1.650_{-0.020}^{+0.020}$ &          $10.70_{-0.30}^{+0.40}$ & $0.154_{-0.005}^{+0.006}$ \\
G072.80-18.72 &  0.143 &        $0.825_{-0.008}^{+0.011}$ &           $5.33_{-0.12}^{+0.17}$ & $0.155_{-0.004}^{+0.005}$ \\
G073.96-27.82 &  0.233 &        $1.710_{-0.020}^{+0.020}$ &          $11.40_{-0.30}^{+0.40}$ & $0.150_{-0.004}^{+0.006}$ \\
G080.38-33.20 &  0.107 &        $0.383_{-0.003}^{+0.003}$ &           $3.66_{-0.06}^{+0.07}$ & $0.105_{-0.002}^{+0.002}$ \\
G080.99-50.90 &    0.3 &        $0.955_{-0.017}^{+0.020}$ &           $6.62_{-0.23}^{+0.32}$ & $0.144_{-0.006}^{+0.008}$ \\
G083.28-31.03 &  0.412 &        $1.240_{-0.020}^{+0.030}$ &           $7.68_{-0.38}^{+0.47}$ & $0.161_{-0.008}^{+0.011}$ \\
G085.99+26.71 &  0.179 &        $0.397_{-0.041}^{+0.047}$ &           $3.31_{-0.57}^{+0.68}$ & $0.120_{-0.024}^{+0.028}$ \\
G086.45+15.29  &   0.26 &        $0.930_{-0.018}^{+0.021}$ &           $5.74_{-0.25}^{+0.33}$ & $0.162_{-0.008}^{+0.010}$ \\
G092.73+73.46 &  0.228 &        $1.070_{-0.020}^{+0.020}$ &           $6.18_{-0.28}^{+0.32}$ & $0.173_{-0.008}^{+0.010}$ \\
G093.91+34.90 &  0.081 &        $0.549_{-0.033}^{+0.044}$ &           $4.68_{-0.50}^{+0.71}$ & $0.117_{-0.014}^{+0.020}$ \\
G096.87+24.21 &    0.3 &        $0.628_{-0.089}^{+0.108}$ &           $5.47_{-0.39}^{+2.40}$ & $0.115_{-0.018}^{+0.054}$ \\
G097.73+38.11 &  0.171 &        $0.687_{-0.016}^{+0.017}$ &           $4.21_{-0.20}^{+0.21}$ & $0.163_{-0.009}^{+0.009}$ \\
G098.95+24.86 &  0.093 &        $0.299_{-0.007}^{+0.014}$ &           $3.25_{-0.15}^{+0.35}$ & $0.092_{-0.005}^{+0.011}$ \\
G106.73-83.22 &  0.292 &        $0.831_{-0.022}^{+0.026}$ &           $5.18_{-0.29}^{+0.36}$ & $0.160_{-0.010}^{+0.012}$ \\
G107.11+65.31 &  0.292 &        $0.694_{-0.028}^{+0.030}$ &           $7.26_{-0.67}^{+0.75}$ & $0.096_{-0.010}^{+0.011}$ \\
G113.82+44.35 &  0.225 &        $0.570_{-0.022}^{+0.028}$ &           $3.87_{-0.21}^{+0.32}$ & $0.147_{-0.010}^{+0.014}$ \\
G124.21-36.48 &  0.197 &        $0.781_{-0.033}^{+0.029}$ &           $5.81_{-0.62}^{+0.57}$ & $0.134_{-0.015}^{+0.014}$ \\
G125.70+53.85 &  0.302 &        $0.964_{-0.043}^{+0.043}$ &           $6.78_{-0.65}^{+0.69}$ & $0.142_{-0.015}^{+0.016}$ \\
G139.19+56.35 &  0.322 &        $1.020_{-0.080}^{+0.140}$ &           $7.42_{-0.68}^{+5.68}$ & $0.137_{-0.017}^{+0.107}$ \\
G149.73+34.69 &  0.182 &        $1.160_{-0.050}^{+0.050}$ &           $7.12_{-0.63}^{+0.71}$ & $0.163_{-0.016}^{+0.018}$ \\
G157.43+30.33 &   0.45 &        $0.964_{-0.031}^{+0.035}$ &           $6.23_{-0.42}^{+0.52}$ & $0.155_{-0.012}^{+0.014}$ \\
G159.85-73.47 &  0.206 &        $0.988_{-0.047}^{+0.041}$ &           $6.73_{-0.70}^{+0.64}$ & $0.147_{-0.017}^{+0.015}$ \\
G164.18-38.89  &  0.074 &        $0.630_{-0.015}^{+0.028}$ &           $5.07_{-0.22}^{+0.45}$ & $0.124_{-0.006}^{+0.012}$ \\
G166.13+43.39 &  0.217 &        $0.868_{-0.019}^{+0.022}$ &           $6.86_{-0.41}^{+0.48}$ & $0.127_{-0.008}^{+0.009}$ \\
G167.65+17.64  &  0.174 &        $0.925_{-0.021}^{+0.026}$ &           $5.88_{-0.30}^{+0.40}$ & $0.157_{-0.009}^{+0.012}$ \\
G171.94-40.65 &   0.27 &        $1.500_{-0.070}^{+0.090}$ &          $12.30_{-2.10}^{+3.40}$ & $0.122_{-0.022}^{+0.034}$ \\
G180.24+21.04 &  0.546 &        $2.340_{-0.050}^{+0.030}$ &          $12.50_{-0.60}^{+0.50}$ & $0.187_{-0.010}^{+0.008}$ \\
G182.44-28.29 &  0.088 &        $0.958_{-0.020}^{+0.011}$ &           $6.93_{-0.18}^{+0.17}$ & $0.138_{-0.005}^{+0.004}$ \\
G182.63+55.82 &  0.206 &        $0.688_{-0.010}^{+0.015}$ &           $4.77_{-0.16}^{+0.26}$ & $0.144_{-0.005}^{+0.008}$ \\
G186.39+37.25 &  0.282 &        $1.240_{-0.050}^{+0.060}$ &           $9.28_{-0.89}^{+1.19}$ & $0.134_{-0.014}^{+0.018}$ \\
G195.62+44.05 &  0.295 &        $0.877_{-0.036}^{+0.049}$ &           $5.82_{-0.49}^{+0.72}$ & $0.151_{-0.014}^{+0.020}$ \\
G195.77-24.30 &  0.203 &        $1.110_{-0.010}^{+0.020}$ &           $6.67_{-0.22}^{+0.32}$ & $0.166_{-0.006}^{+0.009}$ \\
G218.85+35.50 &  0.175 &        $0.430_{-0.019}^{+0.020}$ &           $3.86_{-0.41}^{+0.45}$ & $0.111_{-0.013}^{+0.014}$ \\
G225.92-19.99 &   0.46 &        $2.190_{-0.540}^{+0.280}$ &          $11.00_{-4.40}^{+3.20}$ & $0.199_{-0.094}^{+0.063}$ \\
G226.17-21.91 &  0.099 &        $0.510_{-0.009}^{+0.011}$ &           $4.61_{-0.23}^{+0.32}$ & $0.111_{-0.006}^{+0.008}$ \\
G226.24+76.76 &  0.143 &        $0.834_{-0.005}^{+0.006}$ &           $6.28_{-0.09}^{+0.12}$ & $0.133_{-0.002}^{+0.003}$ \\
G228.15+75.19 &  0.545 &        $1.350_{-0.050}^{+0.070}$ &           $7.94_{-0.60}^{+0.96}$ & $0.170_{-0.014}^{+0.022}$ \\
G228.49+53.12 &  0.143 &        $0.587_{-0.021}^{+0.017}$ &           $5.16_{-0.42}^{+0.35}$ & $0.114_{-0.010}^{+0.008}$ \\
G229.21-17.24 &  0.171 &        $0.656_{-0.068}^{+0.084}$ &           $5.26_{-0.19}^{+1.84}$ & $0.125_{-0.014}^{+0.046}$ \\
G229.94+15.29 &   0.07 &        $0.851_{-0.013}^{+0.012}$ &           $7.01_{-0.26}^{+0.25}$ & $0.121_{-0.005}^{+0.005}$ \\
G236.95-26.67 &  0.148 &        $0.696_{-0.014}^{+0.014}$ &           $5.91_{-0.33}^{+0.32}$ & $0.118_{-0.007}^{+0.007}$ \\
G241.74-30.88 &  0.271 &        $0.759_{-0.022}^{+0.025}$ &           $5.80_{-0.33}^{+0.42}$ & $0.131_{-0.008}^{+0.010}$ \\
G241.77-24.00 &  0.139 &        $0.395_{-0.004}^{+0.004}$ &           $3.29_{-0.06}^{+0.06}$ & $0.120_{-0.003}^{+0.003}$ \\
G241.97+14.85 &  0.169 &        $0.710_{-0.045}^{+0.069}$ &           $4.13_{-0.38}^{+0.64}$ & $0.172_{-0.019}^{+0.031}$ \\
G244.34-32.13 &  0.284 &        $1.140_{-0.030}^{+0.030}$ &           $6.94_{-0.50}^{+0.54}$ & $0.164_{-0.013}^{+0.013}$ \\
G244.69+32.49 &  0.153 &        $0.504_{-0.012}^{+0.017}$ &           $3.70_{-0.19}^{+0.31}$ & $0.136_{-0.008}^{+0.012}$ \\
G247.17-23.32 &  0.152 &        $0.416_{-0.026}^{+0.037}$ &           $3.17_{-0.35}^{+0.54}$ & $0.131_{-0.017}^{+0.025}$ \\
G249.87-39.86 &  0.165 &        $0.402_{-0.028}^{+0.036}$ &           $2.86_{-0.40}^{+0.56}$ & $0.141_{-0.022}^{+0.030}$ \\
G250.90-36.25 &    0.2 &        $0.674_{-0.020}^{+0.027}$ &           $5.36_{-0.36}^{+0.51}$ & $0.126_{-0.009}^{+0.013}$ \\
G252.96-56.05 &  0.075 &        $0.390_{-0.003}^{+0.003}$ &           $3.58_{-0.05}^{+0.04}$ & $0.109_{-0.002}^{+0.001}$ \\
G253.47-33.72 &  0.191 &        $0.571_{-0.033}^{+0.030}$ &           $4.52_{-0.48}^{+0.44}$ & $0.126_{-0.015}^{+0.014}$ \\
G256.45-65.71 &   0.22 &        $0.718_{-0.054}^{+0.057}$ &           $5.42_{-0.76}^{+0.89}$ & $0.132_{-0.021}^{+0.024}$ \\
G257.34-22.18 &  0.203 &        $0.451_{-0.051}^{+0.082}$ &           $3.19_{-0.51}^{+0.88}$ & $0.141_{-0.028}^{+0.047}$ \\
G260.03-63.44 &  0.284 &        $1.000_{-0.020}^{+0.030}$ &           $7.15_{-0.73}^{+0.73}$ & $0.140_{-0.015}^{+0.015}$ \\
G262.25-35.36 &  0.295 &        $1.080_{-0.060}^{+0.070}$ &           $6.59_{-0.70}^{+0.90}$ & $0.164_{-0.020}^{+0.025}$ \\
G262.71-40.91 &   0.42 &        $1.220_{-0.060}^{+0.070}$ &           $9.16_{-0.59}^{+2.11}$ & $0.133_{-0.011}^{+0.032}$ \\
G263.16-23.41 &  0.227 &        $1.120_{-0.010}^{+0.020}$ &           $7.07_{-0.28}^{+0.35}$ & $0.158_{-0.006}^{+0.008}$ \\
G263.66-22.53 &  0.164 &        $1.050_{-0.020}^{+0.020}$ &           $8.59_{-0.45}^{+0.65}$ & $0.122_{-0.007}^{+0.010}$ \\
G266.03-21.25 &  0.296 &        $2.170_{-0.020}^{+0.020}$ &          $12.50_{-0.30}^{+0.30}$ & $0.174_{-0.004}^{+0.004}$ \\
G269.31-49.87 &  0.085 &        $0.319_{-0.016}^{+0.026}$ &           $2.65_{-0.23}^{+0.41}$ & $0.120_{-0.012}^{+0.021}$ \\
G271.19-30.96 &   0.37 &        $1.280_{-0.050}^{+0.040}$ &           $8.38_{-0.53}^{+0.68}$ & $0.153_{-0.011}^{+0.013}$ \\
G271.50-56.55 &    0.3 &        $0.979_{-0.061}^{+0.075}$ &           $8.07_{-0.51}^{+2.11}$ & $0.121_{-0.011}^{+0.033}$ \\
G272.10-40.15 &  0.059 &        $0.779_{-0.005}^{+0.005}$ &           $6.11_{-0.08}^{+0.09}$ & $0.127_{-0.002}^{+0.002}$ \\
G277.75-51.73 &   0.44 &        $1.440_{-0.040}^{+0.050}$ &           $8.96_{-0.59}^{+0.73}$ & $0.161_{-0.011}^{+0.014}$ \\
G278.60+39.17 &  0.307 &        $1.330_{-0.050}^{+0.050}$ &           $9.37_{-0.81}^{+0.87}$ & $0.142_{-0.013}^{+0.014}$ \\
G280.19+47.81 &  0.156 &        $0.744_{-0.047}^{+0.061}$ &           $6.53_{-0.09}^{+1.70}$ & $0.114_{-0.007}^{+0.031}$ \\
G282.49+65.17 &  0.077 &        $0.664_{-0.010}^{+0.011}$ &           $5.25_{-0.20}^{+0.22}$ & $0.126_{-0.005}^{+0.006}$ \\
G283.16-22.93 &   0.45 &        $1.050_{-0.060}^{+0.060}$ &           $7.34_{-0.97}^{+1.14}$ & $0.143_{-0.021}^{+0.024}$ \\
G284.46+52.43 &  0.441 &        $1.690_{-0.030}^{+0.030}$ &          $10.60_{-0.40}^{+0.50}$ & $0.159_{-0.007}^{+0.008}$ \\
G284.99-23.70  &   0.39 &        $1.480_{-0.070}^{+0.080}$ &          $10.10_{-1.20}^{+1.50}$ & $0.147_{-0.019}^{+0.023}$ \\
G285.63-17.24  &   0.35 &        $0.820_{-0.078}^{+0.057}$ &           $6.59_{-0.17}^{+1.00}$ & $0.124_{-0.012}^{+0.021}$ \\
G286.58-31.25 &   0.21 &        $0.718_{-0.015}^{+0.024}$ &           $5.52_{-0.26}^{+0.45}$ & $0.130_{-0.007}^{+0.011}$ \\
G286.99+32.91 &   0.39 &        $2.200_{-0.050}^{+0.060}$ &          $12.20_{-0.70}^{+0.70}$ & $0.180_{-0.011}^{+0.011}$ \\
G288.61-37.65 &  0.127 &        $0.735_{-0.050}^{+0.067}$ &           $4.00_{-0.48}^{+0.70}$ & $0.184_{-0.025}^{+0.036}$ \\
G292.51+21.98 &    0.3 &        $1.130_{-0.020}^{+0.020}$ &           $8.03_{-0.44}^{+0.49}$ & $0.141_{-0.008}^{+0.009}$ \\
G294.66-37.02 &  0.274 &        $0.988_{-0.026}^{+0.036}$ &           $7.20_{-0.39}^{+0.59}$ & $0.137_{-0.008}^{+0.012}$ \\
G304.67-31.66 &  0.193 &        $0.538_{-0.061}^{+0.076}$ &           $4.16_{-0.79}^{+1.10}$ & $0.129_{-0.029}^{+0.039}$ \\
G304.84-41.42 &   0.41 &        $1.140_{-0.030}^{+0.030}$ &           $8.28_{-0.60}^{+0.64}$ & $0.138_{-0.011}^{+0.011}$ \\
G306.68+61.06 &  0.085 &        $0.503_{-0.017}^{+0.007}$ &           $3.93_{-0.31}^{+0.13}$ & $0.128_{-0.011}^{+0.005}$ \\
G306.80+58.60 &  0.085 &        $0.582_{-0.009}^{+0.014}$ &           $4.60_{-0.16}^{+0.26}$ & $0.127_{-0.005}^{+0.008}$ \\
G308.32-20.23  &   0.48 &        $1.050_{-0.040}^{+0.050}$ &           $8.61_{-0.98}^{+1.22}$ & $0.122_{-0.015}^{+0.018}$ \\
G313.36+61.11 &  0.183 &        $1.050_{-0.000}^{+0.000}$ &           $7.87_{-0.09}^{+0.10}$ & $0.133_{-0.002}^{+0.002}$ \\
G313.87-17.10 &  0.153 &        $1.070_{-0.010}^{+0.010}$ &           $8.24_{-0.23}^{+0.25}$ & $0.130_{-0.004}^{+0.004}$ \\
G318.13-29.57 &  0.217 &        $0.743_{-0.038}^{+0.040}$ &           $5.59_{-0.57}^{+0.64}$ & $0.133_{-0.015}^{+0.017}$ \\
G321.96-47.97 &  0.094 &        $0.483_{-0.014}^{+0.017}$ &           $3.95_{-0.23}^{+0.29}$ & $0.122_{-0.008}^{+0.010}$ \\
G324.49-44.97 &  0.095 &        $0.347_{-0.010}^{+0.020}$ &           $3.09_{-0.19}^{+0.40}$ & $0.112_{-0.008}^{+0.016}$ \\
G332.23-46.36 &  0.098 &        $0.706_{-0.006}^{+0.008}$ &           $5.09_{-0.10}^{+0.15}$ & $0.139_{-0.003}^{+0.004}$ \\
G332.88-19.28 &  0.147 &        $0.774_{-0.019}^{+0.021}$ &           $6.22_{-0.33}^{+0.38}$ & $0.124_{-0.007}^{+0.008}$ \\
G335.59-46.46 &  0.076 &        $0.461_{-0.044}^{+0.054}$ &           $3.53_{-0.56}^{+0.74}$ & $0.131_{-0.024}^{+0.031}$ \\
G336.59-55.44 &  0.097 &        $0.488_{-0.036}^{+0.046}$ &           $3.78_{-0.52}^{+0.71}$ & $0.129_{-0.020}^{+0.027}$ \\
G337.09-25.97 &   0.26 &        $0.894_{-0.030}^{+0.034}$ &           $5.75_{-0.41}^{+0.50}$ & $0.155_{-0.012}^{+0.015}$ \\
G342.31-34.90 &  0.232 &        $0.949_{-0.040}^{+0.045}$ &           $6.85_{-0.62}^{+0.74}$ & $0.139_{-0.014}^{+0.016}$ \\
G347.18-27.35 &  0.237 &        $1.100_{-0.050}^{+0.040}$ &           $8.24_{-0.73}^{+0.63}$ & $0.133_{-0.013}^{+0.011}$ \\
G349.46-59.94 &  0.347 &        $2.230_{-0.040}^{+0.040}$ &          $13.50_{-0.60}^{+0.60}$ & $0.165_{-0.008}^{+0.008}$
\label{tab:cluster_sample}
\end{longtable}
\tablefoot{Columns: (1) Name in the ESZ catalog; (2) Cluster redshift; (3) Gas mass; (4) Total mass derived assuming hydrostatic equilibrium; (5) Gas mass fraction. This table is available at \url{http://szcluster-db.ias.u-psud.fr}}
}

\section{Effect of non self-similarity}
\label{annex:B}

The scaling relations between two quantities (X, Y) from \cite{2020ApJ...892..102L} are fitted using the following generic form :
\begin{equation}
    \log\left(\frac{Y}{\mathrm{C1}}\right) = \epsilon + \gamma\log\left(\frac{Z}{\mathrm{C2}}\right) + \delta \log F_z \pm \sigma_{Y|Z}
\end{equation}
with
\begin{equation*}
    \log X = \log Z \pm \sigma_{X|Z},
\end{equation*}
where C1 and C2 are pivot points, $\epsilon$ is the normalisation, $\gamma$ the slope of the relation, $\delta$ its evolution with redshift, $\sigma$ the intrinsic scatter in the two variables and $F_z = E(z)/E(z)^{ref}$ \citep{2020ApJ...892..102L}.
As a result, when we focus on the $M_{HE}-M_{gas}$ relation, we have the following expression (we ignore $\sigma$ for simplicity as they only represent a scatter, i.e. an uncertainty, around a mean value):
\begin{equation}
    \log \left(\frac{M_{HE}}{6\cdot 10^{14}M_\odot}\right) = \epsilon + \gamma\log\left(\frac{M_{gas}}{10^{14}M_\odot}\right)  + \delta \log \left(\frac{E(z)}{E(z)^{ref}}\right)
\end{equation}
\begin{equation}
    \Rightarrow \frac{M_{HE}}{6\cdot 10^{14}M_\odot} = 10^\epsilon \left(\frac{M_{gas}}{10^{14}M_\odot}\right)^\gamma \left(\frac{E(z)}{E(z)^{ref}}\right)^\delta
\end{equation}
In particular, the value for $\delta$ measured by \cite{2020ApJ...892..102L} is $\delta = -0.317 \pm 0.307$. 
Even if this value is consistent with 0 within $1\sigma$, the error bars encompass a large range of values, up to $\delta = -0.6$, which represents a strong evolution. We test therefore the impact of having $\delta \neq 0$.
We find here that this assumption does not affect our results, as the factor $F_z^\delta = (E(z)/E(z)^{ref})^\delta$ does not deviate from 1 when we are close to the reference cosmology. 
In our analysis, the cosmological constraints we find are always in agreement with a \cite{2020A&A...641A...6P} cosmology within $1\sigma$ (see Table \ref{tab:results_varying_v_constant}).
When using this cosmology in our redshift range, we always remain compatible with $F_z^{\delta} = 1$ within $1\sigma$, with less than 1\% of deviation.
This term can thus be ignored in the rest of the analysis, and we are left with the following relation :
\begin{equation}
    \Rightarrow M_{HE}^{1-\gamma} = \frac{6\cdot 10^{14}M_\odot \times 10^\epsilon}{(10^{14}M_\odot)^\gamma} \cdot \left(\frac{M_{gas}}{M_{HE}}\right)^\gamma.
\end{equation}
For displaying purposes we set :
\begin{equation*}
    A = \frac{6\cdot 10^{14}M_\odot \times 10^\epsilon}{(10^{14}M_\odot)^\gamma}.
\end{equation*}
We obtain the following expression, as $M_{HE} = B(M, z)\cdot M_{tot}$ :
\begin{equation}
    \left({B(M, z)\cdot M_{tot}} \right)^{1-\gamma} = A \cdot f_{gas, obs}^\gamma.
\end{equation}
We remind that throughout the paper we assumed a power law evolution of the bias $B(M, z) = B_0 \left(\frac{M_{tot}}{\left<M_{tot}\right>}\right)^\alpha \left(\frac{1+z}{\left<1+z\right>}\right)^\beta$.
We thus have the following relation:
\begin{equation}
    \left(B_0 M_{tot}^\alpha M_{tot} (1+z)^\beta\right)^{1-\gamma} \propto f_{gas,obs}^\gamma
\end{equation}
\begin{equation}
    \Rightarrow f_{gas, obs} \propto M_{tot}^{(\alpha+1)\frac{1-\gamma}{\gamma}}\cdot (1+z)^{\beta\frac{1-\gamma}{\gamma}}
\end{equation}
and the fitted $M_{tot}-M_{gas}$ scaling relation can be summed up as follows:
\begin{equation}
    M_{tot}^{\alpha+1} (1+z)^{\beta(1-\gamma)} \propto M_{gas}^{\gamma}.
\end{equation}
In the paper, we find an evolution of the gas fraction $f_{gas, obs} \propto M_{tot}^{0.057\pm 0.038}(1+z)^{0.64\pm 0.017}$ for the full sample.
There are several possibilities to explain this result when accounting for deviations of the $M_{tot}-M_{gas}$ relation from self-similarity :
\begin{itemize}
    \item The self-similar hypothesis predicts a linear and redshift independent relation between the total mass and the gas mass, i.e. $\alpha + 1 = \gamma$ and $\beta(1-\gamma) = 0$. 
    In this description, all the mass and redshift evolution of the gas fraction comes from the bias.
    From \cite{2020ApJ...892..102L} we have $\gamma = 0.802\pm 0.049$.
    This means that in the self-similar case, we should obtain $\alpha = -0.198\pm 0.049$ i.e. $f_{gas}\propto M_{tot}^{(\alpha+1)\frac{1-\gamma}{\gamma}} = M_{tot}^{0.198\pm 0.049}$,
    which is in tension with our measured value at $2.9\sigma$.
    In addition, from our measured value of $\beta = -0.64\pm 0.17$, we find $\beta(1-\gamma) = -0.13\pm 0.05$, in tension with 0 at $2.6 \sigma$.
    This hints at the fact that the mass and redshift evolution of $f_{gas}$ can not be accounted for solely by an evolution of the bias.
    Instead, a deviation from self-similarity of the $M_{tot} - M_{gas}$ relation is necessary to explain our results.
    \item As we show a deviation of the $M_{tot} - M_{gas}$ relation from self-similarity, one may wonder whether this effect may totally mimic the effect of a mass evolution of the bias. 
    This is however not the case. 
    Indeed, if we set a constant bias $\alpha = \beta = 0$, then $(\alpha+1)\frac{1-\gamma}{\gamma} = \frac{1-\gamma}{\gamma} = 0.057\pm 0.049$ (found with the full sample), leading to $\gamma = 0.946\pm 0.034$, in tension at $4\sigma$ the with the value found by \cite{2020ApJ...892..102L} of $\gamma = 0.802\pm 0.049$.
    Furthermore, putting $\beta = 0$ leads to $\beta \frac{1-\gamma}{\gamma} = 0$, i.e. a redshift independent gas fraction. This is however not what we observe.
    \end{itemize}
    We thus show that the dependence of $f_{gas}$ on mass and redshift cannot be due to deviations from self-similarity or to an evolution of the bias alone. 
     Instead, it can only be explained by a combination of the two effects.

\section{Effect of the prior on the depletion $\Upsilon_0$}
\label{annex:C}

\begin{figure*}
    \centering
    \includegraphics[width=\textwidth, trim = {1cm 1cm 1cm 1cm}, clip]{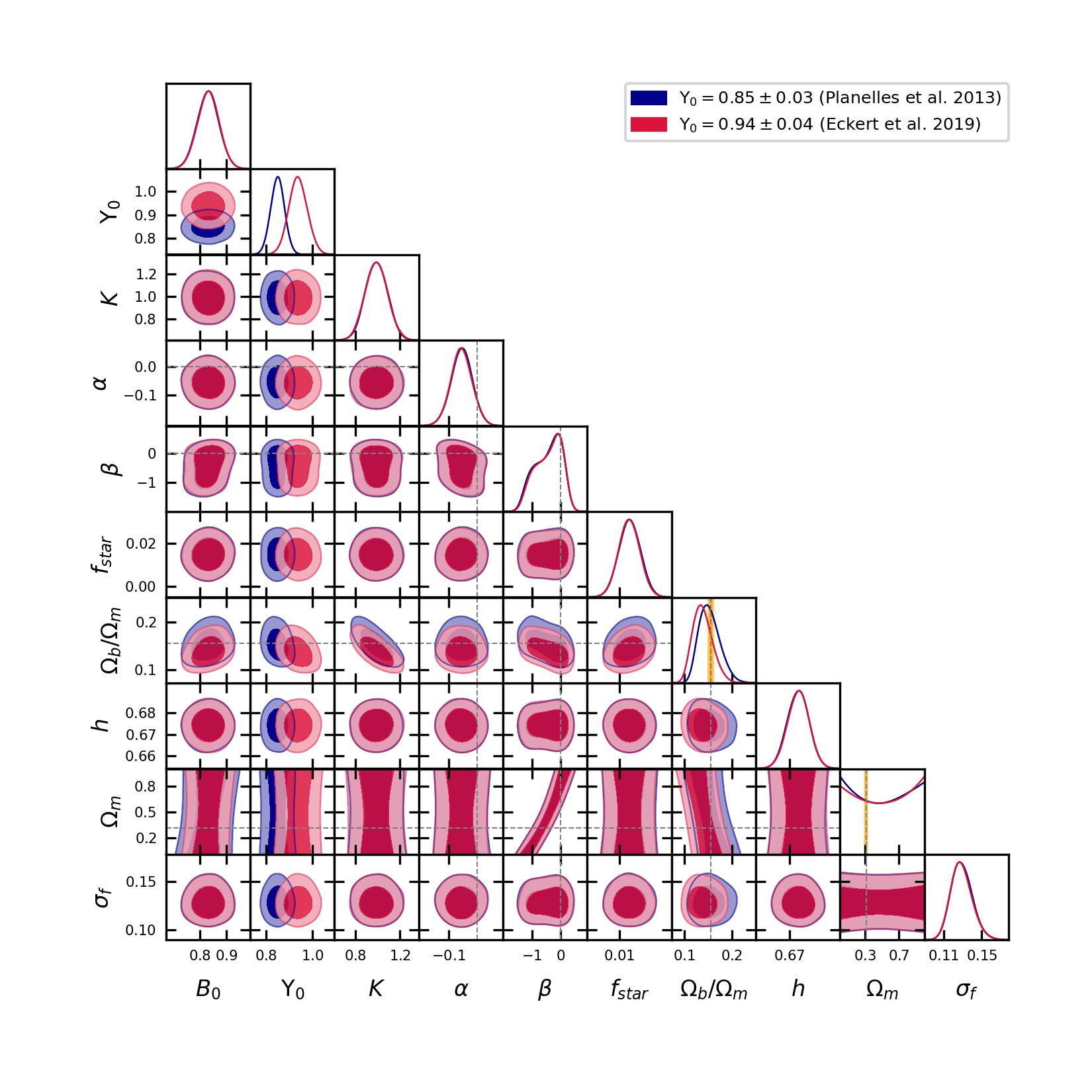}
    \caption{MCMC results for a 10\% shift of the depletion factor in our analysis in the VB scenario. We note that in this case, we assume a prior on the total value of the bias.}
    \label{fig:depletion_check_biasprior}
\end{figure*}

\begin{figure*}
    \centering
    \includegraphics[width=\textwidth, trim = {1cm 1cm 1cm 1cm}, clip]{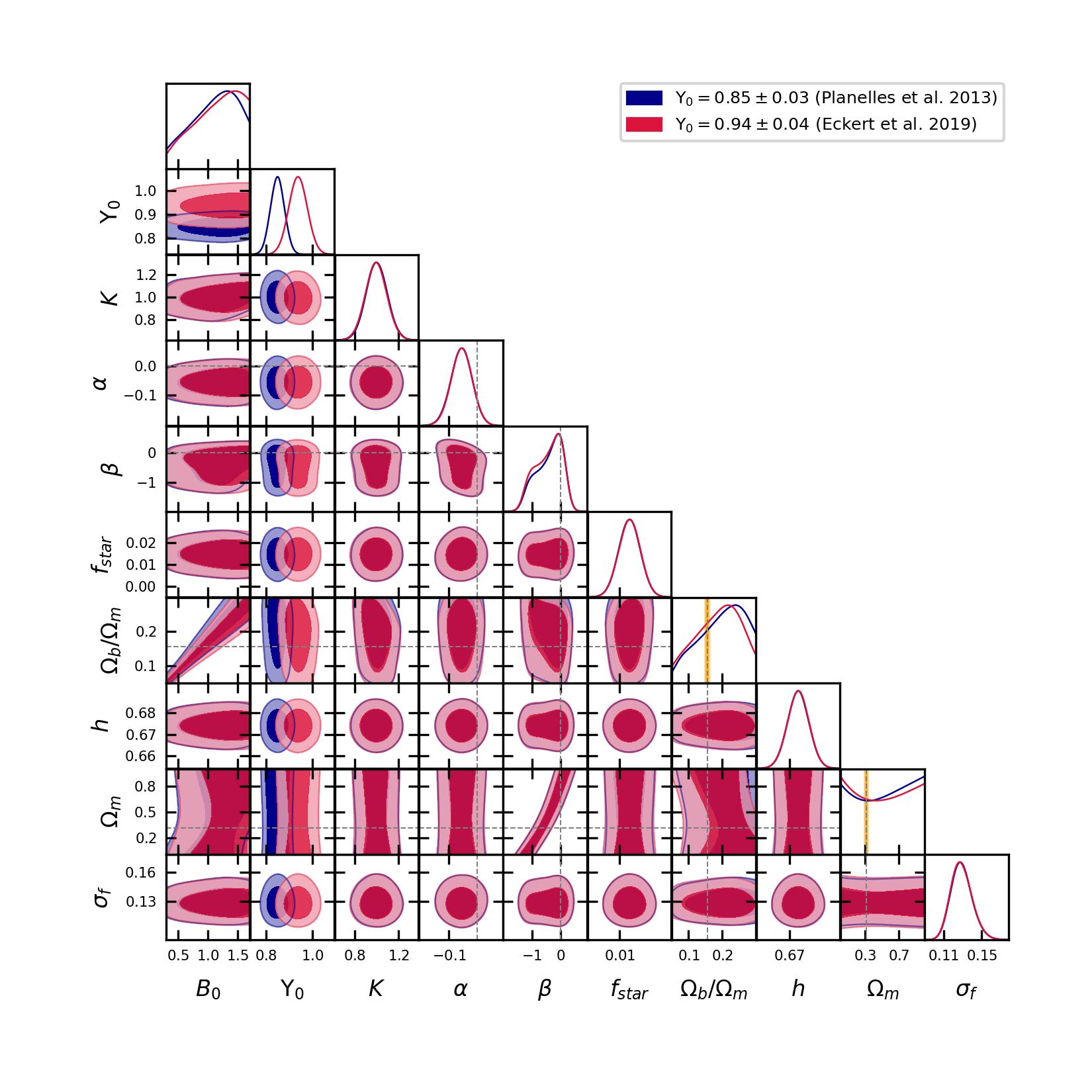}
    \caption{MCMC results for a 10\% shift of the depletion factor in our analysis in the VB scenario, without any prior on the total value of the bias.}
        \label{fig:depletion_check_openbias}
\end{figure*}

\begin{table*}
    \centering
    \caption{Constraints on the model parameters in the VB scenario, after a 10\% shift of the depletion factor, with and without a prior on the total value of the bias. All uncertainties and limits are at the 68\% c.l.}
    \begin{tabular}{lcc|cc}
    \hline \hline
       & {\bf Prior on the total bias} &  & & {\bf No prior on the total bias} \\
        \hline
         & $\Upsilon_0 = 0.85 \pm 0.03$ & $\Upsilon_0 = 0.94 \pm 0.04$ & $\Upsilon_0 = 0.85 \pm 0.03$ & $\Upsilon_0 = 0.94 \pm 0.04$ \\
    \hline 
      $\boldsymbol{B_0}$   & $0.831 \pm 0.04$ & $0.832 \pm 0.04$ & $1.10^{+0.53}_{-0.23}$ & $ > 0.968$ \\
      $\boldsymbol{\alpha}$ & $-0.055 \pm 0.037$ & $-0.056 \pm 0.037$ &  $-0.056 \pm 0.037$ & $ -0.055 \pm 0.037$ \\
      $\boldsymbol{\beta}$ & $-0.43^{+0.60}_{-0.37}$ & $-0.41^{+0.58}_{-0.36}$ & $ -0.40^{+0.59}_{-0.33}$ & $ -0.43^{+0.60}_{-0.36}$ \\
      $\boldsymbol{\Omega_b/\Omega_m}$ & $0.153^{+0.018}_{-0.025}$ & $0.139^{+0.018}_{-0.023}$ & $ 0.198^{+0.095}_{-0.036}$ & $0.188^{+0.082}_{-0.052}$ \\
      $\boldsymbol{\Omega_m}$ & -- & -- & -- & -- \\
      $\boldsymbol{\sigma_f}$ & $0.129^{+0.010}_{-0.012}$ & $0.129^{+0.010}_{-0.012}$ & $ 0.129^{+0.010}_{-0.012}$ & $0.129^{+0.010}_{-0.012}$ \\
      $\boldsymbol{(\Upsilon_0, f_*, h)}$ & {\it = prior} & {\it = prior} & {\it = prior} & {\it = prior} \\
      \hline 
    \end{tabular}
    \label{tab:depletion_check}
\end{table*}

In order to properly assess the effect of the depletion factor on our conclusions, we performed runs of the VB scenario when considering two different values of the depletion, the one we assumed throughout the paper $\Upsilon_0 = 0.85 \pm 0.03$ from \cite{2013MNRAS.431.1487P} and $\Upsilon_0 = 0.94 \pm 0.04$ from \cite{2019A&A...621A..40E}.
    
When we apply a 10\% shift in the value of the depletion, we do not observe a significant shift in the value of the other parameters, as we show in Figure \ref{fig:depletion_check_biasprior} and Table \ref{tab:depletion_check}. 
We however acknowledge that this result is due to the presence of a prior on the total value of the bias, which we assume in the first part of the analysis.
To try to quantify the impact of this prior, we leave the bias fully free in second MCMC runs.
We show the results in Figure \ref{fig:depletion_check_openbias} Table \ref{tab:depletion_check}. 
Similarly to the previous case, we do not observe a particular shift of the constraints from a shift of the depletion, mostly because the posterior distribution are extremely wide.
In particular, we do not find a degeneracy between $\Upsilon$ and other parameters, including $B_0$ or cosmological parameters.
We thus claim that although the depletion factor is indeed a strong systematic in our analysis, we address it properly. 
\end{appendix}

\end{document}